\documentclass{jpp}
\usepackage{graphicx}

 \usepackage[utf8]{inputenc}
 \usepackage[T1]{fontenc}
 \usepackage{amsmath}
 \usepackage{empheq}
 \usepackage{hyperref}

 \usepackage{xcolor}
 \usepackage{soul} 
 \definecolor{red}{rgb}{1.0,0.0,0.0}
 \definecolor{gre}{rgb}{0.5,0.5,1.5}
 \definecolor{blu}{rgb}{0.0,0.0,1.0}

\graphicspath{{./IllustrationsTheory/}{./IllustrationsResults/}{./Illustrations/}}  

 \shorttitle{Application of Lagrangian techniques}
 \shortauthor{S. Guinchard, W. Sengupta and S. R. Hudson}

 \title{Application of Lagrangian techniques for calculating the on-axis rotational transform}

 \author{S. Guinchard\aff{1}
  \corresp{\email{salomon.guinchard99@gmail.com}},
  W. Sengupta\aff{1,2}
 \and S. R. Hudson\aff{1}}

\affiliation{\aff{1}Princeton Plasma Physics Laboratory,
Princeton, NJ, 08540, USA
\aff{2}Department of Astrophysical Sciences, Princeton University, Princeton, NJ, 08543, USA}

 \input{head/commands.sty}

\begin{document}

 \maketitle

\begin{abstract}
The Floquet exponents of periodic field lines are studied through the variations of the magnetic action on the magnetic axis, which is assumed to be elliptical. 
The near-axis formalism developed by Mercier, Solov'ev and Shafranov is combined with a Lagrangian approach.
The on-axis Floquet exponent is shown to coincide with the on-axis rotational transform, and this is a coordinate-independent result.
A discrete solution suitable for numerical implementation is introduced, which gives the Floquet exponents as solutions to an eigenvalue problem. This discrete formalism expresses the exponents as the eigenvalues of a $6\times 6$ matrix. 
\end{abstract}


\section{The need for and the origin of the rotational transform}


Magnetostatic equilibria are characterized by the following equations, 
\beq
\grad p = \bJ \times \bB, \;\;
\rot \bB = \bJ, \;\;
\Div \bB = 0,
\label{eq:MHS_integrable}
\eeq
where $\bB$ is the magnetic field, $\bJ$ is the current density, and $p$ is the scalar pressure. 
These equations imply that magnetic field lines lie on surfaces of constant $p$, and that the constant pressure surfaces are toroidal \citep{KruskalKulsrud}.
The phase portrait of $\bB$, where the magnetic field lines are treated like integral curves of a Hamiltonian dynamical system, is characterized by the topology of the level sets of $p$.
A (regular) \emph{flux surface} is a closed surface along which the pressure is constant and $\grad p\neq 0$. An equilibrium field satisfying Eq.(\ref{eq:MHS_integrable}) such that $\grad p\neq0$ almost-everywhere is said to be \emph{integrable} \citep{BurbyDuignanMeiss}. 
A magnetic field line action can be defined \citep{CaryLittlejohn}, which serves as a starting point for the Lagrangian integration carried out in this paper.
 
For magnetic confinement of plasmas in toroidal geometries, that the magnetic field lines rotate poloidally (the short way) as they rotate toroidally (the long way) around the torus is essential for canceling charged particle drifts, which would otherwise lead to loss \citep{SpitzerStell}.
The number of poloidal rotations that a field line achieves per toroidal period is called the rotational transform, $\iota$ \citep{SpitzerStell}. 

The Poincaré first \emph{return map} is defined as follows: given a toroidal magnetic configuration in the domain $\Omega$, and any poloidal section $\Sigma$ of $\Omega$, the first return map is defined by the intersection of field lines and $\Sigma$ after one toroidal period. 
By Brouwer's fixed-point theorem \citep{Brouwer}, there must be at least one \emph{magnetic axis}, defined as a closed field line such that its intersection with any poloidal plane $\Sigma$ is a fixed point of this map \citep{MercierLuc}. 
In the axisymmetric case, such an axis is a circle located at the center of a family of nested flux surfaces.

Magnetic fields with continuously nested flux surfaces, equivalently invariant surfaces of Hamiltonian dynamical flows, are only guaranteed if there is a continuous symmetry \citep{Noether1918}. 
Nonetheless, many surfaces with sufficiently irrational rotational transform will persist under small symmetry-destroying perturbations \citep[see][]{Kolmogorov_1954, Arnold_63, Moser_62} and also the texts and reviews \citep{Moser_73,Arnold_78,Meiss_Rev,Lichtenberg_Lieberman_92}. 

In this work, only the following needs to be assumed: there is at least one surface that is invariant under the Poincaré map, and everywhere inside this surface the toroidal component of the field is nonzero.


Except at fixed points, iterating the Poincaré map once will rotate points around the magnetic axis, as measured by mean of an arbitrary angle, $\theta$. 
Following a field line by successive iterations of the map gives a sequence of angles $\theta_n$, measured between iterations $n$ and $n+1$.
This sequence of angular displacements enables to define the rotational transform, 
\beq
\iota := \lim_{N\uparrow \infty}\frac{1}{2\pi N}\sum_n^N\theta_n. \label{eq:iota_no_coord}
\eeq
This is purely geometric, in the sense that it describes the linking of a field line around a reference axis. 
Denoting the toroidal and poloidal magnetic fluxes, $\psi$ and $\chi$ respectively, then, as shown by \citet{MercierLuc}, $\iota$ can also be expressed as the ratio of differential fluxes,
\beq
\iota =  \frac{d\chi}{d\psi}.\label{eq:iota_fluxes}
\eeq
Mercier \citep[see][]{Mercier_1964, helander_2014} expressed the rotational transform as an integral along the axis 
\beq
\iota = N + \frac{1}{2\pi } \oint \frac{d\ell}{\cosh{\eta}} \Big( \frac{J_0}{2B_0}+\delta' - \tau\Big) ,\label{eqn:Mercier_rotational_transform}
\eeq
with $\ell$ denoting the arc length. The on-axis current density and magnetic field are denoted by $J_0$ and $B_0$ respectively, and the torsion of the axis by $\tau$. 
The eccentricity of flux surfaces is described by $\eta$, and $\delta$ is a parameter describing their rotation  around the axis, with  $\delta' := d\delta/d\ell$. 
$N$ is an integer coming from the phase of the rotation term $\delta$ \citep[see][]{Pfefferlé}.
(Note: \citet{SpitzerStell} had identified independently that a way to generate some rotational transform is to give torsion to an axis.) 
 
This expression 
provides invaluable insight, showing how rotational transform can be produced by plasma currents, 
as is used by tokamaks,  
or by geometrical shaping, as is used by stellarators. 
Mercier's expression 
was derived starting from the fluxes definition Eq.(\ref{eq:iota_fluxes}). 
The fluxes were expanded in a power series of the distance to  the axis by constructing custom polar coordinates, so-called Mercier coordinates, which are described in Appendix \ref{appendix:near-axis-coordinates}.

In this paper, we present a derivation of Mercier's formula using Lagrangian integration. 
This approach expresses the rotational transform as a Floquet exponent of the field lines.
In Section \ref{sec:MagAction}, the magnetic field line action is defined. Stationary curves of the action are shown to be field lines, enabling to identify a magnetic axis. Assumption is made in this paper that the axis is elliptical, but the formalism can be applied to the hyperbolic case. In Section \ref{sac:Verif_Mercier}, Mercier's formula, Eq.(\ref{eqn:Mercier_rotational_transform}) is derived from the second variation of the action. 
The result is obtained through a near-axis expansion of the null eigenspace of the second variation operator. 
The periodicity enables the use of a Floquet description of the solutions. This same result is derived through the theory of the Hill's infinite determinant, analogous to Schrödinger's equation in a periodic potential. In both cases, the rotational transform is shown to be a Floquet exponent of the null eigenspace of the second variation operator. In Section \ref{sec:discrete_action}, a discrete formalism is introduced so that the Floquet exponents can be solved for numerically.

\section{Description of the magnetic field line action}\label{sec:MagAction}

Let us start by defining the magnetic field line action, as introduced by \cite{CaryLittlejohn}. 
This is defined as a line integral depending on the integration contour. In this paper, those will be assumed to be closed curves, embedded in a toroidal domain $\Omega$, in which there exists a magnetic field with non-vanishing toroidal component, and at least one smooth magnetic surface enclosing the curve. Let us consider a closed, differentiable curve $\mathcal{C} \subset \mathbb{R}^3$ with total length $L$, closing after one toroidal transit. The latter assumption that the curve closes after one toroidal period is a necessary condition for a magnetic axis. 
$\cC$ is parameterized by the $C^1$, $L-$periodic vector-valued function
\beq
\begin{split}
\bx:[0,L]&\rightarrow \mathbb{R}^3\\
\ell &\mapsto \bx(\ell)\in \cC,
\end{split}
\eeq 
$\ell$ being the arc length and we note $\bx' := d\bx/d\ell$. 
The action is defined as the circulation along $\mathcal{C}$ of the magnetic vector potential $\bA$, with $\bB = \curl \bA$ in $\Omega$,
\beq
\mathcal{S} := \oint_{\mathcal{C}}d\ell \; \bA \boldsymbol{\cdot} \bx'.
\label{eqn:action}
\eeq
Properties of the magnetic field are accessible through calculus of variations, performed on the action. The variations are performed with respect to changes in the geometry of the curve. 
The first variation with respect to a variation $\delta \bx$ is
\beq
\delta \mathcal{S}[\delta \bx] = \oint_{\mathcal{C}} d\ell\; \bx^\prime \times {\bB} \boldsymbol{\cdot}  \delta \bx,
\label{eqn:actionvariation}
\eeq
which shows that stationary curves are tangential to the magnetic field and hence field lines. We used that $\delta \bA [\delta \bx]=\delta \bx \boldsymbol{\cdot}\grad \bA$. 
From Eq.(\ref{eqn:actionvariation}), a magnetic axis can be identified. We will focus in particular on elliptical axes, which can be shown to be local minima of the action, but one can apply the following formalism to hyperbolic axes as well.
Let us denote an elliptical axis by $\cC_a$. Additional properties of the field appear at higher orders of variations of $\mathcal{S}$. For the rotational transform in particular, the second order variation $\delta^2 \cS$ is of interest. 


\section{Verification of Mercier's formula for the Floquet exponent}\label{sac:Verif_Mercier}

\subsection{The second variation as an operator}

In order to express the on-axis rotational transform, the second order variation of the action applied to the axis needs to be derived. 
Assume that an elliptical axis $\cC_a$ has been found as a stationary curve of the action. The formalism is identical for hyperbolic axes, but in this paper, we focus on the elliptic case. The second variation of the action performed from the axis $\cC_a$ is 
\beq
\delta^2 \mathcal{S}[\delta \bx] = \oint_{\mathcal{C}_a} d\ell\;\delta(\bx^\prime \times {\bB} \boldsymbol{\cdot} \delta{\bx}) =  \oint_{\mathcal{C}_a} d\ell\;\delta \bx \boldsymbol{\cdot} (\delta \bx^\prime \times \bB + \bx' \times \delta \bB),
\eeq
where $\boldsymbol{f}':=d\boldsymbol{f}/d\ell$ for any $\boldsymbol{f}$. Using $\delta \bB = \delta \bx \boldsymbol{\cdot} \nabla \bB$ and the Einstein's summation convention, we write the second variation as an operator,
\beq
    \delta^2 \mathcal{S} = \oint_{\cC_a} d\ell\; \delta \bx^i \frac{\delta^2 S}{\delta \bx^i \delta \bx^j} \delta \bx^j ,
\eeq
where
\begin{align}
    \frac{\delta^2 \mathcal{S}}{\delta \bx^i \delta \bx^j} = \epsilon_{ijk}\bB^k\frac{d}{d\ell} + \epsilon_{imk}\bx'^m \partial_j \bB^k,
\end{align}
and $i,j,k,m \in \{1,2,3\}$, which in matrix form reads
\begin{align}
    \mathsfbi{M} :=  \frac{\delta^2 \mathcal{S}}{\delta \bx \delta \bx}= -\left(\mathsfbi{I}\times \bB \right) \frac{d}{d\ell}+ \bx' \times (\nabla \bB)^\mathsf{T}.
    \label{eq:TensorM}
\end{align}
The identity tensor is denoted by $\mathsfbi{I}$. Note that the covariant tensor $\mathsfbi{M}$ can be easily symmetrized  \citep[see][]{HUDSON20094409}. However, the direction in which the derivative $d/d\ell$ is taken has to be chosen carefully. We shall continue with the non-symmetric form Eq.\eqref{eq:TensorM}.

We will show that the null eigenspace of $ \mathsfbi{M}$ is of particular interest. Indeed, by periodicity of the field lines considered to compute the action, the eigenspaces of the operator $\mathsfbi{M}$ are related to geometric properties of the neighboring field lines, including the rotational transform. 
Let $\bv$ be a null eigenfunction of $\mathsfbi{M}$ such that
\begin{align}
    \mathsfbi{M}\bv =\mathbf{0}.
    \label{eq:null_eigenv}
\end{align}
For $\bv$ to be nontrivial, Eq.(\ref{eq:null_eigenv}) rewrites
\begin{align}
    \det\left( \mathsfbi{M} \right) =0.
    \label{eq:det_M_0}
\end{align}
We will demonstrate that the condition Eq.\eqref{eq:null_eigenv} is satisfied by solutions $\boldsymbol{v}$, such that their Floquet exponent is the on-axis rotational transform, and Eq.(\ref{eq:det_M_0}) will serve in the discrete formalism introduced in Section \ref{sec:discrete_action}. 
To solve Eq.\eqref{eq:null_eigenv}, the near-axis formalism developed by Mercier, Solov'ev and Shafranov \citep{PRL5,MercierLuc} will be used. Near-axis formalism in the inverse coordinate approach, where the flux surface $\psi$ is used as a coordinate, has proved to be very successful in understanding quasisymmetry \cite{garren_Boozer_1991existence,landreman_sengupta2018direct,landreman_sengupta2019_higher_order,rodriguez_sengupta_bhattacharjee2023constructing}. However, for this work, the near-axis formalism in direct or Mercier-Solov'ev-Shafranov coordinates \citep{jorge_sengupta_landreman2020near,jorge_sengupta_landreman2020construction,sengupta_rodriguez_jorge_landreman_2024stellarator} is more relevant.

\subsection{Derivation of the Floquet exponent from the second variation}

The operator is expanded in the Solov'ev-Shafranov's coordinates, introduced in Appendix \ref{Sol_Shaf_appendix}. The latter set of coordinates is closely related to the Mercier coordinates, presented in Appendix \ref{Appendix:mercier}. 

In what follows, it will be assumed that the magnetic axis is a closed curve $\cC_a\subset \mathbb{R}^3$ parameterized by the vector field $\br_0$, with the arc length $\ell$ as parameter, and a total length $L$.  
The basis of expansion is taken to be $\{\boldsymbol{e}_1, \boldsymbol{e}_2, \boldsymbol{e}_3\}$, whose expressions in terms of $\bt$, $\bcN$ and $\bcB$ are given in Appendix \ref{Sol_Shaf_appendix}, but the results will always be expressed in terms of the Solov'ev-Shafranov's vectors $\{\bt, \bcN, \bcB\}$. 
Additionally, the following notation is adopted: $f'=df/d\ell$ for any $f$. 

\subsubsection{Magnetic field expansion near an elliptic magnetic axis}

From Eq.(\ref{eq:TensorM}), the magnetic field $\bB$ needs to be described in the vicinity of the axis $\br_0$ for $\mathsfbi{M}$ to be expanded in the near-axis formalism. 
The expansion of $\bB$ in Solov'ev-Shafranov's coordinates is carried in the limit $x,y\ll 1$, or in other words, is limited to linear terms only. 
The magnetic field can be written in contravariant form as 
\beq
\sqrt{|g|}\bB= \sqrt{|g|}B^1 \boldsymbol{e}_1+\sqrt{|g|}B^2 \boldsymbol{e}_2+\sqrt{|g|}B^3 \boldsymbol{e}_3 ,
\label{eq:Bcontravariant}
\eeq
with 
\beq
    \sqrt{|g|}B^1= a_1 x+a_2 y , \quad \sqrt{|g|}B^2= b_1 x+b_2 y, \quad \sqrt{|g|}B^3= B_0+c_1 x+c_2 y,
\eeq
where $a_1,a_2,b_1,b_2, c_1,c_2$ are periodic functions of $\ell$, $B_0$ the 0th order magnitude of $\bB$ and $\sqrt{|g|} = h$ as defined in Appendix (\ref{Sol_Shaf_appendix}). 
Therefore, in the limit $x,y\ll1$, $h=1$ and the magnetic field reads
\beq
\begin{split}
    \bB =  \left(B_0 +c_1 x+c_2 y\right)\bt  + \left[a_1 x+(a_2 +u' B_0 )y\right]&\boldsymbol{\mathcal{N}}\\
    +\left[(b_1-u' B_0) x+b_2 y\right]&\boldsymbol{\mathcal{B}}+ \boldsymbol{\mathcal{O}}(x^2+y^2).
    \label{eq:B_rep_Sol_Shaf}
\end{split}
\eeq
\noindent A direct calculation shows that,
\beq
\begin{split}
    \nabla \bB=\;&\kappa B_0 \bt \bn +a_1 \bcN\bcN+b_2 \bcB\bcB+B_0' \bt\bt + (a_2+u' B_0) \bcB\bcN \\
    &+(b_1-u' B_0)\bcN\bcB + (c_1 \boldsymbol{\mathcal{N}}+c_2 \boldsymbol{\mathcal{B}})\boldsymbol{t},\\
    \boldsymbol{x'}\times  (\nabla \boldsymbol{B})^\mathsf{T}
    =\;&\kappa B_0 \bb \bt +\boldsymbol{\mathcal{B}} \left[a_1 \boldsymbol{\mathcal{N}}+(a_2 +u' B_0) \boldsymbol{\mathcal{B}} \right]-\boldsymbol{\mathcal{N}} \left[(b_1-u' B_0) \boldsymbol{\mathcal{N}}+b_2 \boldsymbol{\mathcal{B}} \right],
    \label{eq:grad_vec_B_sol_shaf}
    \end{split}
\eeq
which yields, using dyadic algebra
\beq
\begin{split}
     \Div\bB = \mathsfbi{I}\boldsymbol{:}\nabla \bB =&\; B_0' + a_1+b_2,\\
    \curl \bB = -\mathsfbi{I}\boldsymbol{\overset{\times}{.}}\nabla \bB =&\; (b_1-a_2-2 u' B_0) \bt+  (c_2 -\kappa B_0 \sin\delta )\bcN  \\
    &- (c_1-\kappa B_0 \cos \delta)\bcB.
\end{split}
\eeq
Here $\kappa$ denotes the local curvature of the axis $\br_0$. Additionally, Eq.\eqref{eq:cal_N_B} has been used to rewrite $\bb$ in terms of $\bcN$ and $\bcB$. 
Requiring that $\nabla \boldsymbol{\cdot} \bB=0$ and $\nabla \times \bB=J_0 \bt$, one finds
\begin{align}
    a_1+b_2= -B_0', \quad \frac{b_1-a_2}{2B_0}=u'+\frac{J_0}{2B_0}, \quad c_1 =\kappa B_0 \cos \delta,  \quad c_2 =\kappa B_0 \sin \delta.
    \label{eq:constriants_from_div_curl_B_sol_shaf}
\end{align}
So far, only two constraints have been derived for the four expansion functions $a_1,a_2,b_1,b_2$. 
Turning to the Mercier representation, Eq.\eqref{eq:form_B} and Eq.\eqref{eq:B1x_exp} of $\bB$ brings in additional conditions. Together with the definitions of  $\bcN$ and $\bcB$,
\beq
\begin{split}
    \frac{\bB}{B_0} =&\quad  \bt\left[1+\kappa \sqrt{x^2+y^2} \cos{(u-\delta)}\right] \\
                                          &+ \bcN \left[ -\frac{1}{2}\left(\frac{B_0'}{B_0}+\eta'\right)x + \left(\frac{-J_0}{2 B_0}+ \frac{\frac{J_0}{2B_0}- \tau +\delta'}{ \cosh{\eta}}
\sinh{\eta}\right) y\right]\\
    									  &+ \bcB \left[ \left(\frac{J_0}{2 B_0}+ \frac{\frac{J_0}{2B_0}- \tau +\delta'}{ \cosh{\eta}}
\sinh{\eta}\right) x-\frac{1}{2}\left(\frac{B_0'}{B_0}-\eta'\right)y \right]\\
& + \boldsymbol{\mathcal{O}}(x^2+y^2),
    \label{eq:B_rep_Mercier}
\end{split}
\eeq
where $\delta$, as explained in Appendix \ref{appendix:near-axis-coordinates}, describes the rotation of elliptical flux surfaces around the expansion axis $\br_0$. As for $\eta$, it describes the eccentricity of the flux surfaces around $\br_0$. Defining 
\beq
    \Omega_0(\ell)=\frac{\frac{J_0}{2B_0}- \tau +\delta'}{ \cosh{\eta}},\label{eq:Omega0}
\eeq 
the comparison between the Mercier representation Eq.\eqref{eq:B_rep_Mercier} and the Solov'ev-Shafranov representation Eq.\eqref{eq:B_rep_Sol_Shaf} yields for the expansion coefficients
\beq
\begin{split}
    \frac{a_1}{B_0}&=-\frac{1}{2}\left(\frac{B_0'}{B_0}+\eta'\right), \quad  \frac{a_2}{B_0}=-\left(u'+\frac{J_0/2}{ B_0}\right)+\Omega_0\sinh{\eta}, \\
    \frac{b_2}{B_0}&=-\frac{1}{2}\left(\frac{B_0'}{B_0}-\eta'\right), \quad \frac{b_1}{B_0}=\left(u'+\frac{J_0/2}{ B_0}\right)+\Omega_0\sinh{\eta},
    \label{eq:a1a2b1b2_exp}
\end{split}
\eeq
which satisfy the relations Eq.\eqref{eq:constriants_from_div_curl_B_sol_shaf}. 
From the definition of $\Omega_0$, we can further simplify $a_2$ and $b_1$ as
\begin{align}
    \frac{a_2}{B_0}=-\Omega_0 e^{-\eta}, \quad \frac{b_1}{B_0}=\Omega_0 e^{\eta}.
\end{align}
As a check of correctness for the above expressions,  one can show that $\bB\boldsymbol{\cdot} \nabla \psi=0$, where 
\beq
\begin{split}
   \sqrt{g} \bB\boldsymbol{\cdot} \nabla &= \sqrt{g}B^1\partial_x +\sqrt{g}B^2\partial_y+\sqrt{g}B^3\partial_\ell \\
   				       &= B_0\partial_\ell + (a_1 x + a_2 y)\partial_x + (b_1 x + b_2 y)\partial_y.
\end{split}
\eeq
The flux surface function $\psi$ for the elliptic case is given in the Mercier representation by Eq.\eqref{eq:psi_exp_Mercier} or equivalently in the Solov'ev-Shafranov representation by
\begin{align}
\psi(x,y)=B_0(e^{\eta}x^2 + e^{-\eta}y^2).
    \label{eq:psi_exp_Sol_Shaf}
\end{align}

\subsubsection{The second variation tensor and its null eigenspace}

Now that the magnetic field has been properly expressed and expanded in Solov'ev-Shafranov's coordinates, the obtained $\bB$ and $\grad \bB$ can be substituted  in Eq.\eqref{eq:TensorM} to expand the second variation tensor. From Eq.\eqref{eq:grad_vec_B_sol_shaf} we find that it is given by
\beq
\begin{split}
    \mathsfbi{M}       :=& \; \mathsfbi{M_1} \frac{d}{d\ell}+\mathsfbi{M_2},  \\
    \mathsfbi{M_1}   =& \;(\boldsymbol{\mathcal{N}} \boldsymbol{\mathcal{B}}-\boldsymbol{\mathcal{B}} \boldsymbol{\mathcal{N}})B_0, \\
    \mathsfbi{M_2}   =&\; \kappa B_0 \left[\boldsymbol{\mathcal{B}}\cos{\delta} - \boldsymbol{\mathcal{N}}\sin{\delta} \right]\bt +\boldsymbol{\mathcal{B}} \left[a_1 \boldsymbol{\mathcal{N}}+(a_2+u' B_0) \boldsymbol{\mathcal{B}} \right]\\
    & \;-\boldsymbol{\mathcal{N}} \left[(b_1-u' B_0) \boldsymbol{\mathcal{N}}+b_2 \boldsymbol{\mathcal{B}} \right].
\end{split}\label{eq:tensor_M_sol_shaf}
\eeq
Observing the dyadic form of $\mathsfbi{M}$, it is clear that tangential components $v^t$ of of the eigenvectors $\bv=(v^t, v^{\cN}, v^{\cB})^{\mathsf{T}}$ will not contribute to $ \mathsfbi{M}\bv =\boldsymbol{0}$. 
Therefore, $v^t$ can be absorbed by redefining  $v^{\cN}$ and $v^{\cB}$. 
Thus, the null eigenvectors are chosen to be of the form  $\bv= v^{\cN} \boldsymbol{\mathcal{N}}+ v^{\cB} \boldsymbol{\mathcal{B}}$. We note that
\begin{align}
    \frac{d}{d\ell}\bv = \left(v^{\cN\prime}+u' v^{\cB}\right)\bcN + \left(v^{\cB\prime}-u' v^{\cN}\right)\bcB +\kappa\left[v^{\cN} \cos{(u-\delta)} + v^{\cB} \sin{(u-\delta)}\right]\bt.
\end{align}
The tangential terms $\propto \bt$ are not relevant since they do not contribute to $\mathsfbi{M} \boldsymbol{v} =\boldsymbol{0}$. The equations for $v^{\cN},v^{\cB}$ then read
\begin{align}
    B_0 
    \begin{pmatrix}
     \;\;0 & +1 \;\; \\-1 & 0 
    \end{pmatrix}
    \frac{d}{d \ell} 
    \begin{pmatrix}
        v^{\cN}\\ v^{\cB}
    \end{pmatrix} +
    \begin{pmatrix}
       -b_1& & -b_2\\ a_1 & & a_2 
    \end{pmatrix}
     \begin{pmatrix}
        v^{\cN}\\ v^{\cB}
    \end{pmatrix} =0.
    \label{eq:Mdotv_matrix_form_sol_shaf}
\end{align}
From Eq.\eqref{eq:Mdotv_matrix_form_sol_shaf}, by periodicity of the $a_i/B_0$ and $b_i/B_0$, $i=1,2$, the system can immediately be rewritten in the form of a periodic system 
\beq
\frac{d\bv}{d\ell}=\mathsfbi{A}(\ell) \bv, \quad \bv = \left(\begin{array}{c}v^{\cN} \\v^{\cB}\end{array}\right). \label{eq:periodic-syst-Sol-Shaf}
\eeq
Here, $\mathsfbi{A}(\ell)$ is periodic in $\ell$. 
The Floquet theorem, Eq.(\ref{eq:Floquet_thm}), can be applied to conclude that the solution must be of the form
\begin{align}
   \bv = \mathsfbi{U}(\ell)e^{\mathsfbi{C} \ell/L},
    \label{eq:Floquet_solution}
\end{align}
\noindent where $\mathsfbi{U}$ is a symplectic periodic matrix (with period $L$) and $\mathsfbi{C}$ is a constant Hamiltonian matrix \citep{Duignan_2021}. The eigenvalues of $\mathsfbi{C}$ are the Floquet exponents, which must be purely imaginary near an elliptic axis and real in the hyperbolic case.

Although the system is already in a form that allows to solve for the exponents $\nu$ by mean of Eq.\eqref{eq:Floquet_solution}, identifying the matrices $\mathsfbi{C}$ and $\mathsfbi{U}$ can be somewhat troublesome. For that reason, Eq.\eqref{eq:Mdotv_matrix_form_sol_shaf} be alternatively rewritten in the following ordinary differential equation (ODE) form
\begin{align}
    \frac{dv^{\cN}}{a_1 v^{\cN} + a_2 v^{\cB}}= \frac{dv^{\cB}}{b_1 v^{\cN} + b_2 v^{\cB}} = \frac{d \ell}{B_0},
    \label{eq:v_characteristic_eqn}
\end{align}
which shows that the eigenvector $\bv$ satisfies the general characteristic equation along the magnetic field 
\begin{align}
    \frac{d\cX}{\sqrt{g}B^1(\cX,\cY,\ell)} = \frac{d\cY}{\sqrt{g}B^2(\cX,\cY,\ell)}= \frac{d\ell}{\sqrt{g}B^3(\cX,\cY,\ell)}.
     \label{eq:B_characteristic_eqn}
\end{align}
The eigenvector $\bv$ components $v^{\cN},v^{\cB}$ can be identified with $\cX,\cY$, the displacements of the magnetic field line from the closed field line $\br_0$ along the rotated normal and binormal directions. The expansion coordinates $(x,y)$ ought not to be confused with $(\cX,\cY)$, which are the solutions of the characteristic ODEs Eq.\eqref{eq:B_characteristic_eqn}. 
The equations for $(\cX,\cY)$ are
\beq
\begin{split}
    &\cX'+\frac{1}{2}\left(\frac{B_0'}{B_0}+\eta'\right)\cX + \Omega_0 e^{-\eta}\cY=0,\\
    &\cY'+\frac{1}{2}\left(\frac{B_0'}{B_0}-\eta'\right)\cY - \Omega_0 e^{+\eta}\cX=0.
    \label{eq:cXcY_ODEs}
\end{split}
\eeq
Introducing the variables $X,Y$ defined through
\beq
   \cX= \frac{1}{\sqrt{B_0}}e^{-\eta/2}X(\ell), \quad  \cY= \frac{1}{\sqrt{B_0}}e^{+\eta/2}Y(\ell),\label{eq:oscill_harm_SolShaf}
\eeq
which is possible since the periodicity is conserved, the system reduces to the simple harmonic oscillator, 
\begin{align}
    X' + \Omega_0 Y =0, \quad Y'-\Omega_0 X =0, \quad \Omega_0 = \frac{\frac{J_0/2}{B_0}- \tau +\delta'}{ \cosh{\eta}},
    \label{eq:X_Y_system_Solov_shafr}
\end{align}
with `time-dependent' frequency $\Omega_0(\ell)$,  $\ell$ being the time-like parameter.
Using the complex variable $Z=X+i Y$, the system reshapes as a single complex ODE,
\begin{align}
    Z'-i \Omega_0 Z =0 \quad \Rightarrow \quad Z(\ell)= Z_0\exp{\int_0^\ell i\Omega_0(s) ds}.
\end{align}
Separating the periodic and non-periodic parts of the exponential, we obtain
\begin{align}
    Z(\ell) = Z_p(\ell) e^{i \nu \ell/L}, \quad Z_p(\ell)= Z_0 \exp{\int_0^\ell i\widetilde{\Omega}_0(s) ds}, \label{eq:Z_expr_Solov_shafr}\\
    \overline{\Omega} := \frac{1}{L}\int_0^L \Omega_0(s) ds, \quad \widetilde{\Omega}_0= \Omega_0 - \overline{\Omega}.
    \nonumber
\end{align}

\noindent Comparing Eq.\eqref{eq:Z_expr_Solov_shafr} with the Floquet form Eq.\eqref{eq:Floquet_solution}, we find that $\nu$ is given by 
\begin{align}
\nu = \int_0^L \Omega_0(s) ds \pmod{2\pi} = \oint \frac{\frac{J_0(s)/2}{B_0(s)}- \tau(s) +\delta'(s)}{ \cosh{\eta(s)}} ds \pmod{2\pi} = 2\pi \iota.\label{eq:floquet_exp=rotational}
\end{align}
It matches the expression for the rotational transform Eq.(\ref{eqn:Mercier_rotational_transform}) up to a factor $2\pi$, as $\nu$ represents an angle and $\iota$ the number of turns that this angle constitutes.  It follows that solving for the null eigenspace of the second variation tensor expanded in the Solov'ev-Shafranov near-axis formalism yields the correct on-axis rotational transform. The exact same approach can be followed in the hyperbolic case. 

\subsubsection{Derivation of the Floquet exponent from the Hill's determinant equation}
Here we shall pursue an alternative approach to obtain the Floquet exponent, using the theory of the Hill's infinite determinant. 
This way of solving for the Floquet exponents of a periodic system has been known for a long time \citep{magnus1953infinite}; however, to the authors' knowledge, it is the first time that such an approach is used from the magnetic action. 
We start with the system Eq.\eqref{eq:cXcY_ODEs} written as
\begin{align}
    \cX'=\frac{1}{B_0}(a_1 \cX +a_2 \cY), \quad  \cY'=\frac{1}{B_0}(b_1 \cX +b_2 \cY).
    \label{eq:cXcYeqns}
\end{align}
Eliminating $\cY$ from Eq.\eqref{eq:cXcYeqns}, we obtain the following second order ODE for $\cX$,
\begin{align}
    \cX'' +2 C_1 \cX + C_2 =0,
    \label{eq:2nd_order_cX}
\end{align}
where
\begin{align}
    2C_1 = -\frac{a_2'}{a_2}+2\frac{B_0'}{B_0}, \quad C_2 = \cD -\frac{a_2}{B_0}\left(\frac{a_1}{a_2}\right)', \quad \cD = \frac{1}{B_0^2}(a_1 b_2 -a_2 b_1).
    \label{eq:C1C2D_exp}
\end{align}
We can eliminate the first derivative term from Eq.\eqref{eq:2nd_order_cX} by the change of variables
\begin{align}
    \cX = \exp{\left(-\int C_1 d\ell\right)}\Psi = \frac{\sqrt{a_2}}{B_0}X,
    \label{eq:xtoX_transform}
\end{align}
which leads to
\begin{align}
\Psi'' + \omega^2 \Psi=0, \quad \omega^2 \equiv C_2 -C_1'-C_1^2.
    \label{eq:Hills_eqn_Psi}
\end{align}
Eq.\eqref{eq:Hills_eqn_Psi} is in the form of Hill's equation or a Schrodinger equation with a periodic potential $\omega^2$. 
We note that the linear transformation Eq.\eqref{eq:xtoX_transform} implies that both $\cX$ and $\Psi$ have the same Floquet exponent. 
This is because the multiplication factor $\sqrt{a_2}/B_0$ is periodic in nature and therefore does not change the Floquet exponent.

Leveraging the periodicity of $\omega^2$, $\omega^2(\ell+L)=\omega^2(\ell)$, we can Fourier expand
\beq
\omega^2(\ell)=\sum_{k\in \mathbb{Z}}\Omega_ke^{i\ell\frac{2\pi}{L}k},\label{eq:Fourier_omega2}
\eeq
where the $\{\Omega_k\}_{k\in \mathbb{Z}}$ denote the Fourier coefficients of $\omega^2$. Let the fundamental solutions of Eq.\eqref{eq:Hills_eqn_Psi} be given by $\Psi_{\pm}(\ell)$ such that
\begin{align}
    \Psi_{+}(0)=1, \quad  \Psi_{-}(0)=0, \qquad \Psi'_{+}(0)=0, \quad  \Psi'_{-}(0)=1,
    \label{eq:Fund_soln_Psi}
\end{align}
where these conditions have been chosen such that the basis functions $\Psi_{\pm}$ are linearly independent and with unit Wronskian. 
The Floquet solutions are given by
\beq
\begin{split}
    \Psi_{+}(\ell)&= e^{+i \nu \ell}\sigma_{+}(\ell), \quad \sigma_{+}(\ell+L)=\sigma_{+}(\ell), \quad \sigma_+(0)=1,\\
    \Psi_{-}(\ell)&= e^{-i \nu \ell}\sigma_{-}(\ell), \quad \sigma_{-}(\ell+L)=\sigma_{-}(\ell), \quad \sigma_-(0)=0,
\end{split}
\eeq
where $\sigma_{+}$ and $\sigma_{-}$ satisfy the periodicity condition and the linear independence, and $\nu$ denotes the Floquet exponents that we seek for. Nothing more needs to be known about the functions $\sigma_{\pm}$ since their coefficients will not appear in the final expression for the Floquet exponents. Let us now Fourier expand the Floquet solution as
\beq
    \Psi_+ =  e^{i \nu \ell}\sum_{n\in \mathbb{Z}}b_n e^{i\frac{2\pi n }{L}\ell} = \sum_{n\in \mathbb{Z}} b_n e^{i \ell \left(\frac{2\pi n }{L}+ \nu\right)},
\eeq
where $\{b_n\}_{n\in \mathbb{Z}}$ is the set of Fourier coefficients of $\sigma_{+}$. 
The Fourier expansion of Eq.(\ref{eq:Hills_eqn_Psi}) reads
\beq
\begin{split}
\Psi_+^{''}(\ell)+\omega^2(\ell)\Psi_+(\ell)=&\;-\sum_{n\in \mathbb{Z}}b_n \left( \nu + \frac{2\pi n}{L}\right)^2 e^{i \ell \left(\nu + \frac{2n\pi}{L}\right)}\\
								&\;+ \left(\sum_{k\in \mathbb{Z}}\Omega_ke^{i\ell\frac{2\pi}{L}k} \right)\left(\sum_{n\in \mathbb{Z}} b_n e^{i \ell \left(\frac{2\pi n }{L}+ \nu\right)}\right)\\
								=&\;\sum_{n\in \mathbb{Z}}\left(\sum_{k}\Omega_kb_{n-k}-\left(\frac{2\pi n }{L}+ \nu\right)^2b_n\right)e^{i\ell\left(\frac{2\pi n }{L}+ \nu\right)}\\
								=&\;0.
\end{split}\label{eq:Hills_fourier_expanded}
\eeq
In order for Eq.(\ref{eq:Hills_fourier_expanded}) to be verified, the following condition has to hold, 
\beq
\sum_{k\in \mathbb{Z}}\Omega_kb_{n-k}-\left(\frac{2\pi n }{L}+ \nu\right)^2b_n=0,\quad n \in \mathbb{Z}, \label{eq:system_Hill_Fourier}
\eeq 
which can be rewritten in the following matrix form, after dividing Eq.(\ref{eq:system_Hill_Fourier}) by $\Omega_0-(2\pi n/L+\nu)^2$:
\beq
\sum_{m\in \mathbb{Z}}B_{nm}b_m=0, \quad n \in \mathbb{Z},
\eeq
where the matrix $B$ is defined as 
\beq
B_{nn}=1, \quad B_{nm}=\frac{\Omega_{n-m}}{\Omega_0-\left(\frac{2\pi n}{L}+\nu\right)^2}.
\eeq
It yields the following determinant equation,
\beq
\det{B_{nm}} = 0.\label{eq:det_B=0}
\eeq
The above determinant can be considered a function of the Floquet exponents $\nu$,
\beq
\det{B_{nm}(\nu)}:=\Delta(\nu)\equiv \det{\left(\delta_{nm} + \frac{\Omega_{n-m}}{\Omega_0-\left(\frac{2\pi n}{L}+\nu\right)^2}\right)}, \quad n,m \in \mathbb{Z}.
\eeq
For Eq.\eqref{eq:det_B=0} to be verified, we seek for the Floquet exponents $\nu$ such that $\Delta(\nu)=0$. Following \cite{SpecialFunctions}, we are now going to show that the exponents can be deduced from the very simple expression Eq.(\ref{eq:characteristic_exp_from_sin}). One notes that $\Delta(\nu)$ is a $2\pi$-periodic function, with poles in $\nu = \pm \sqrt{\Omega_0} - 2\pi n/L$, $n \in \mathbb{Z}$. It can be shown that the determinant is absolutely convergent in the whole $\nu$-plane, expect in these poles. Therefore, $\Delta$ is meromorphic. 
Moreover, as the imaginary part $\Imag{(\nu)}\rightarrow \pm \infty$, $\Delta(\nu)\rightarrow 1$. Let us now define the complex function $f$ as 
\beq
f(\nu):= \cot{\frac{L}{2}(\nu - \sqrt{\Omega_0})} - \cot{\frac{L}{2}(\nu + \sqrt{\Omega_0})}.
\eeq
It is useful to introduce $f$ since it has the same poles as $\Delta$ and the periodicity. Moreover, it is bounded as $\Imag{(\nu)}\rightarrow \pm \infty$. This way, there has to exist a constant $K\in \mathbb{C}$ such that 
\beq
D(\nu) \equiv \Delta(\nu) + Kf(\nu)
\eeq
has no singularity in the whole $\nu$ plane. Together with the fact that it is bounded as $|\nu|\rightarrow \infty$, according to Liouville's theorem, $D$ is a constant function. In the limit $|\nu|\rightarrow \infty$, we see that $D=1$. To determine $K$, take $\nu=0$,
\beq
\nu =0 \implies K = \frac{1-\Delta(0)}{f(0)} = \frac{1-\Delta(0)}{2\cot{\frac{\sqrt{\Omega_0}L}{2}}}.\label{eq:constant_K}
\eeq
Using the value of $K$ from Eq.\eqref{eq:constant_K}, the Floquet exponent equation reduces to 
\beq
\sin^2{\nu\frac{L}{2}} = \Delta(0)\sin^2{\frac{\sqrt{\Omega_0}L}{2}}.\label{eq:characteristic_exp_from_sin}
\eeq
Although Eq.\eqref{eq:characteristic_exp_from_sin} is very simple, one obstacle remains to compute the exponents: one has to evaluate the infinite determinant, $\Delta(0)$. 
We refer to \cite{SpecialFunctions} for some approximations of $\Delta(0)$. For instance, when the $\Omega_n$ are sufficiently small, $\Delta(0)$ can be approximated by the order-3 determinant with $B_{00}$ as central element (we remind that the determinant involves summation over all $\mathbb{Z}$), providing
\beq
\Delta(0)\simeq 1+\frac{2 \Omega_1^2}{\Omega_0\left( 4 - \Omega_0\right)^2}+\frac{2 \Omega_1^2 \Omega_2}{\Omega_0\left( 4 - \Omega_0\right)^2}-\frac{\Omega_2^2}{\left( 4 - \Omega_0\right)^2}.
\eeq 
However, this approximation breaks down when the coefficients become too large as $n$ increases. As of the exponents computed, they might not all be suitable for $\iota$. One has to discard the results that are not relevant. Moreover, we emphasize that the exponents may be shifted by $2k \pi/L$, with $k$ integer, without changing the mathematics of the system, by periodicity. However, one has to carefully choose the adapted value for $\iota$, by setting the appropriate phase shift (see Eq.(\ref{eqn:Mercier_rotational_transform})).

In a nutshell, the second variation of the magnetic action, at an elliptical axis, has been expanded in the Solov'ev-Shafranov near-axis formalism. It has been shown that the null-eigenspace of this operator yields the correct on-axis rotational transform, upon applying Floquet theory to solve for the latter. On the other hand, solving for the null-eigenspace has led to a system that could be rewritten in the form of a Hill's equation. The Floquet exponents appear as solutions of an infinite determinant equation, that can only be approximated analytically. The derivation of the on-axis rotational transform as a Floquet exponent of the null-eigenvectors of $\delta^2\cS$ has also been carried on in Mercier coordinates, to support the previous conclusions - see  Appendix (\ref{app:mercier_derivation}).

Magnetic confinement devices' design needs fast and accurate computation of $\iota$.  
Currently, the most widely used method to compute $\iota$ is to use field-line tracing \citep{FieldLineTracing}. 
Field-line tracing methods compute $\iota$ as an infinite time limit of following a field line. 
In practice, it can be done by following a field line sufficiently long to achieve convergence. 
As an alternative, one can also compute Greene’s residue of the axis, which is related to $\iota$ \citep[see][]{Greene_stochastic,Hanson_Cary_84,Hudson_rational_Surfaces}. 
Even though it still involves O.D.E. integration, we can limit ourselves to just one circuit around the torus.

In the following section, we introduce a new method to compute $\iota$ as a solution to a discrete problem involving the magnetic action. The action is discretized in a similar way as in \citet{MACKAY198392} and \citet{hudson_suzuki}.

 \section{Discrete formalism: piecewise action}\label{sec:discrete_action}

We have seen that the problem described in the previous section, involving the second order variation of the magnetic action to determine the on-axis Floquet exponent, leads to the operator equation Eq.\eqref{eq:null_eigenv}. 
In another approach, magnetic axis can be discretized, and the rotational transform can be determined from the multipliers of the latter curve.
The multipliers have been shown to be linked to the residue of the curve \citep[see][]{MACKAY198392, Greene_stochastic}. 
This discrete approach is explored in what follows. 
We consider
\beq
\begin{split}
\mathcal{S} &= \sum_{i=1}^{n-1} \int_{\mathcal{C}_i} \bA \boldsymbol{\cdot} d\boldsymbol{\ell},\\
&= \sum_{i=1}^{n-1} S(\bx_{i},\bx_{i+1}),
\end{split}\label{eq:discrete_action}
\eeq
meaning the curve has been discretized with $n\in \mathbb{N}$ points. So far, the way the $\cC_i$ are defined is not important. Only the endpoints matter. The particular case where the $\cC_i$ are segments is given in Appendix \ref{app:piecewise}. The following notation will be adopted for the derivatives,
\beq
\begin{split}
\bS_1^{[i,i+1]}&:= \nabla_{\bx_i} S^{[i,i+1]}=\nabla_{\bx_i} S(\bx_i, \bx_{i+1}),\\
\bS_2^{[i,i+1]} &:= \nabla_{\bx_{i+1}} S^{[i,i+1]}=\nabla_{\bx_{i+1}} S(\bx_i, \bx_{i+1}),
\end{split}
\eeq
as well as for the second-order derivatives,
\beq
\mathsfbi{S}_{21}^{[i,i+1]} := \nabla_{\bx_{i+1}}\bS_1^{[i,i+1]},
\eeq
and similarly for $\mathsfbi{S}_{12}$, $\mathsfbi{S}_{11}$ and $\mathsfbi{S}_{22}$. Let us also define generalized periodic orbits of type $(q)$, as orbits with 
\beq
\bx_{i+q} = \bx_i,
\eeq
for some $q\in \mathbb{N}$. Therefore, the magnetic axis as discretized above is a general periodic orbit of type $(n)$.
The terminology \emph{orbit} for field lines is justified by the Hamiltonian behavior of the magnetic field. 
Moreover, the discretized action Eq.(\ref{eq:discrete_action}) satisfies the periodicity condition
\beq
\sum_{i=1}^{n-1} S(\bx_{i+n},\bx_{i+n+1}) = \sum_{i=1}^{n-1} S(\bx_{i},\bx_{i+1}) + C,
\eeq
where the Calabi invariant $C=0$.  In the one-dimensional case, as dealt with in \cite{MACKAY198392}, the existence of extrema of the action is ensured by an additional convexity condition on the Lagrangian, $-L_{12}>0$. 
In our three-dimensional case, this can be generalized in that the second order derivatives tensors $\mathsfbi{S}_{12}^{[i,i+1]}$ ought to be negative definite for all $i$,
\beq
\bx^{\mathsf{T}}{\mathsfbi{S}_{12}}^{[i,i+1]}\bx <0, \quad \bx \in \mathbb{R}^{*3}, \quad 1\leq i \leq n.\label{eq:generalized_convex}
\eeq
This way, the action of periodic orbits of type $(n)$ is bounded below and Poincaré-Birkhoff theorem \citep{PoincareBirkhoff} ensures the existence of at least two stationary trajectories among the space of all generalized periodic paths of type $(n)$, one that minimizes the action and is elliptic, and one that is a saddle called \emph{minimax} and is usually hyperbolic but can be alternating hyperbolic, where hyperbolic and elliptic describe the behavior of nearby trajectories.

For an $(n)$-periodic curve that extremizes the action, the latter has to be stationary with respect to an arbitrary variation in its geometry $\delta \bx_i$, and the stationarity condition can be expressed in terms of the previously defined derivatives,
\beq
\begin{split}
 \delta S [\delta \bx_i] = \Big[\nabla_{\bx_i} S^{[i-1,i]}+ \nabla_{\bx_i} S^{[i,i+1]} \Big]\cdot \delta \bx_i&=0,\\
\Leftrightarrow   \bS_2^{[i-1,i]} + \bS_1^{[i,i+1]} &= \mathbf{0}.\label{eq:stationarity}
\end{split}
\eeq
Note that Eq.\eqref{eq:stationarity} holds for any $1\leq i \leq n$, and the stationarity condition is expressed for each point in terms of the two nearest neighbors. This way, a magnetic axis can be identified. Similarly to the continuous case dealt with in Section \ref{sac:Verif_Mercier}, the neighboring field lines have to satisfy  
\beq
 \nabla_{\bx_{i-1}} \bS_2^{[i-1,i]} \cdot \delta \bx_{i-1} + \nabla_{\bx_{i+1}}\bS_1^{[i,i+1]} \cdot \delta \bx_{i+1}\\
+ \nabla_{\bx_{i}}\Big(\bS_2^{[i-1,i]} + \bS_1^{[i,i+1]}\Big) \cdot \delta \bx_{i}
= \mathbf{0}, \label{eq:tangent_discretised_orbits}
\eeq
 which we rewrite in terms of the arguments of the action as
\beq
\begin{split}
\mathsfbi{S}_{12}^{[i-1,i]}\cdot \delta \bx_{i-1}  + \mathsfbi{S}_{21}^{[i,i+1]}\cdot \delta \bx_{i+1}  + \Big( \mathsfbi{S}_{22}^{[i-1,i]} + \mathsfbi{S}_{11}^{[i,i+1]}\Big)\cdot \delta \bx_{i} = \mathbf{0}, \quad  1\leq i\leq n.\\
\end{split}\label{Stationarity_Tensor}
\eeq
From \cite{MACKAY198392}, we know that the multipliers $\lambda$ of a ($n$)-periodic orbit are defined by existence of a tangent orbit satisfying
\beq
\delta \bx_{i+n} = \lambda \delta \bx_i,
\eeq
so the following holds
\beq
\delta \bx_{n+1} = \lambda \delta \bx_{1}, \quad \delta \bx_0 = \lambda^{-1}\delta \bx_n.\label{eq:multipliers_periodic_curve}
\eeq
They can be written in their Floquet form \citep{Greene_stochastic},
\beq
\lambda = e^{i\nu},
\eeq where the Floquet exponent $\nu$ describes the average rotation angle per period of the ($n$)-orbit, therefore, the rotational transform $\iota$. Since Eq.(\ref{Stationarity_Tensor}) is valid for any $1\leq i\leq n$, together with Eq.(\ref{eq:multipliers_periodic_curve}), it can be rewritten in the tensor form,
\beq
\begin{split}
&\left(\begin{array}{cccc}\left( \mathsfbi{S}_{22}^{[01]} + \mathsfbi{S}_{11}^{[12]} \right) &  \mathsfbi{S}_{12}^{[12]} & \hspace{0.1cm} & \lambda^{-1}\mathsfbi{S}_{21}^{[01]} \\ 
\mathsfbi{S}_{21}^{[12]} & \mathsfbi{S}_{12}^{[23]} & \ddots & \hspace{0.1cm} \\ 
 &  \hspace{-3mm}\ddots & \hspace{2mm}\ddots  & \mathsfbi{S}_{12}^{[n-1,n]} \\ 
\lambda\mathsfbi{S}_{12}^{[n,n+1]} &   & \hspace{-2mm}\mathsfbi{S}_{21}^{[n-1,n]} & \left( \mathsfbi{S}_{22}^{[n-1,n]} + \mathsfbi{S}_{11}^{[n,n+1]} \right)\end{array}\right)
\left(\begin{array}{c}  \delta \bx_{1}\\ \vdots \\  \\ \vdots \\ \delta \bx_{n} \end{array}\right) = \mathbf{0},
\end{split}\label{tensor_expre}
\eeq
where the block-tridiagonal form with corners arises naturally. Denoting by $\mathsfbi{M}$ the matrix of second derivatives, we get an equation for the multipliers:
\beq
\mathsfbi{M}(\lambda)\delta \bx = \mathbf{0}.\label{MLambda}
\eeq
The blank spaces in $\mathsfbi{M}$ are blocks of $0$. Note that each block in $\mathsfbi{M}$ is of size $3\times 3$ since the $\mathsfbi{S}_{lm}$ contain the second derivatives of a line integral embedded in three-dimensional space. 
Thus, for a $(n)$-periodic curve, $\mathsfbi{M}\in \mathcal{M}_{3n\times 3n}$, and $\delta \bx \in \mathbb{R}^{3n}$.  For Eq.(\ref{MLambda}) to hold, the determinant of $\mathsfbi{M}$ ought to be zero to avoid the trivial solution $\delta \bx = \mathbf{0}$.

For a block tridiagonal matrix defined as in Eq.(\ref{tensor_expre}) with $\lambda \in \mathbb{C}$, an expression exists to compute the determinant \citep{Molinari_2008},
\beq
\begin{split}
\det \mathsfbi{M}(\lambda) &= \frac{(-1)^{3n}}{(-\lambda)^3}\det\Big(\mathsfbi{T}_S-\lambda \mathsfbi{I}_{6} \Big)\det\Big( \prod_{i=1}^{n}\mathsfbi{S}_{12}[i,i+1]\Big),\\
\mathsfbi{T}_S &= \prod_{i=1}^{n}\left(\begin{array}{cc} -{\mathsfbi{S}^{-1}_{12}}^{[i,i+1]}(\mathsfbi{S}_{22}^{[i-1,i]}+\mathsfbi{S}_{11}^{[i,i+1]}) & \hspace{0.2cm}-{\mathsfbi{S}_{12}^{-1}}^{[i,i+1]}\mathsfbi{S}_{12}^{[i-1,i]}  \\ \mathsfbi{I}_{3} & \hspace{0.2cm} \mathsfbi{0}\end{array}\right),\\
\end{split}\label{eq:eigenvalues_transfer}
\eeq
which required to define the so-called transfer matrix $\mathsfbi{T}_S$. Our generalized convexity condition Eq.(\ref{eq:generalized_convex}) that ensured existence of minimizing orbits garanties that the determinant of the transfer matrix can be computed as $\mathsfbi{S}_{12}^{[i,i+1]}$ is invertible $\forall i$. 
 The mutlipliers $\lambda$ are then given by the solutions of
\beq
\lambda^{-3}  \det\Big(\mathsfbi{T}_S-\lambda \mathsfbi{I} \Big) = 0\label{multipliers},
\eeq
so they are the non-zero eigenvalues of $\mathsfbi{T}_S$. 
Once the multipliers have been computed, the Floquet exponent can be easily determined from Eq.(\ref{eq:floquet_exp=rotational}).

Alternatively, note that Eq.(\ref{tensor_expre}) gives a recursive relation for the $\delta\bx_i$,
\beq
\begin{split}
\left(\mathsfbi{S}_{22}^{[01]} + \mathsfbi{S}_{11}^{[12]} \right) \delta \bx_1 + \mathsfbi{S}_{12}^{[12]}\delta\bx_2 + \lambda^{-1}\mathsfbi{S}_{21}^{[01]}\delta\bx_n &=0,\\
\lambda\mathsfbi{S}_{12}^{[n,n+1]}\delta\bx_1 +  \mathsfbi{S}_{21}^{[n-1,n]}\delta\bx_{n-1} + \left(\mathsfbi{S}_{22}^{[n-1,n]} + \mathsfbi{S}_{11}^{[n,n+1]} \right) \delta \bx_n &= 0,\\
\mathsfbi{S}_{21}^{[k-1,k]}\delta\bx_{k-1} + \left(\mathsfbi{S}_{22}^{[k-1,k]} + \mathsfbi{S}_{11}^{[k,k+1]} \right) \delta \bx_k + \mathsfbi{S}_{12}^{[k, k+1]}\bx_{k+1} &=0;\;  2\leq k\leq n-1.
\end{split}\label{eq:recursive_var_eq}
\eeq
Eq.(\ref{eq:recursive_var_eq}) can be rewritten as 
\beq
\left(\begin{array}{c} \delta \bx_{n+1} \\ \delta \bx_{n} \end{array}\right)  = \mathsfbi{J} \left(\begin{array}{c} \delta \bx_{1} \\ \delta \bx_{0}\end{array}\right),
\eeq
with 
\beq
\mathsfbi{J} := \prod_{k=1}^{n}\left(\begin{array}{cc} -{\mathsfbi{S}^{-1}_{12}}^{[k,k+1]}(\mathsfbi{S}_{22}^{[k-1,k]}+\mathsfbi{S}_{11}^{[k,k+1]}) & \hspace{0.2cm}-{\mathsfbi{S}_{12}^{-1}}^{[k,k+1]}\mathsfbi{S}_{12}^{[k-1,k]}  \\ \mathsfbi{I} &  \mathsfbi{0}\end{array}\right). 
\eeq
The multipliers of a periodic orbit are the eigenvalues of the derivatives of the return map around the orbit, $\mathsfbi{J}$ \citep{MACKAY198392}, so upon comparison with Eq.(\ref{eq:eigenvalues_transfer}), $\mathsfbi{J}=\mathsfbi{T}_S$ confirming the result from Eq.(\ref{multipliers}).

The strength of this method resides in the fact that the problem is reduced to finding the eigenvalues of a $6\times6$ matrix. In fact, the size of the operator matrix $\mathsfbi{M}$ increases linearly with the number of discretization points, but of interest are solely the eigenvalues of the matrix $\mathsfbi{T}_S$, whose size is $6\times6$, no matter how many discretization points have been used. The derivation of the aforementioned results in the case where the curve of interest is discretized by piece-wise linears is given in Appendix \ref{app:piecewise}.

\section{Conclusion}\label{sec:conclusion}

 In this paper, after having introduced the rotational transform, the Hamiltonian behavior of toroidal magnetic fields was used to motivate the definition of a magnetic action. 
 The latter action served as starting point to express the on-axis rotational transform from a novel approach. 
 The focus has been made on elliptical magnetic axes, but this method applies to hyperbolic axes as well, and more generally, to any periodic field line. 
 
 The action and resulting properties were studied through the lens of the calculus of variations, where variations of the curves' geometry were performed. 
 The first variation led to the result that extremizing curves are magnetic field lines, enabling to identify a magnetic axis. 
 The nature of the axis then defines whether the curve is a local minimum or a saddle point. Such a characteristic can be derived looking at the second order variation. 
 The second variation also sheds light on the geometry of the neighboring field lines. Studying the null-eigenspace of the second variation enabled the derivation of the on-axis rotational transform, with a focus made on elliptical axes. The second variation, seen as an operator, was expanded in the near-axis formalism developed by Mercier, Solov'ev and Shafranov, to yield a system of periodic differential equations. The periodicity of the system allowed the use of Floquet theory to solve for the eigenspace. The key result is that the Floquet exponents of the axis were shown to match Mercier's expression Eq.\eqref{eqn:Mercier_rotational_transform} of the rotational transform. Additionally, solving for the null-eigenspace was shown to lead to a Hill's equation, from which the Floquet exponents were expressed as solutions of an infinite determinant equation.  

Following this continuous derivation of the on-axis rotational transform through the Floquet exponents of the null-eigenspace of the second variation, a discrete approach was introduced. 
It consists in discretizing the field line of interest and by linearity, to define the action as a sum of piecewise actions. 
This approach was described by \citet{MACKAY198392} for one-dimensional Lagrangian systems. We provide a generalization of this method as our action is based on field lines who are in essence three-dimensional. 
Solving for the Floquet exponents was shown to be closely related to solving for the multipliers of the curve, described by \citet{MACKAY198392} and \citet{Greene_stochastic} and the parallel between the two approaches was made as a consistency check. 
The on-axis Floquet multiplier is computed by finding the eigenvalues of a $6\times 6$ matrix, made of second derivatives of the action. 
Its efficiency in comparison with field line tracing will be studied in future work.

Finally, this paper applies the near-axis formalism to recovering the rotational transform from a Lagrangian approach. 
Although the method has been introduced for elliptical axes, also called $O-$points, it can be applied to hyperbolic axes, or $X-$points, with the only difference being that the Floquet exponents are real, not purely imaginary; and, more generally, to all periodic orbits. 
The discrete approach stands as an alternative to compute the rotational transform instead of following field lines. We aim at using this method in stellarator optimization. Lagrangian integration has recently been applied to determine the sensitivity of the geometry of the magnetic axis to perturbations \citep{Hudson_coils}.

\section*{Acknowledgements}
We dedicate this paper to Bob Dewar. The authors would like to thank A. Bhattacharjee and E. J. Paul for stimulating discussions and helpful suggestions.

\section*{Funding}

W.S. was supported by a grant from the Simons Foundation/SFARI (560651, AB), and the Department of Energy Award No. DE-SC0024548.
This manuscript is based upon work supported by the U.S. Department of Energy, Office of Science, Office of Fusion Energy Sciences, and has been authored by Princeton University under Contract Number DE-AC02-09CH11466 with the U.S. Department of Energy.

\section*{Declaration of interests}
The authors report no conflict of interest.

\section*{Data availability}
The data that support the findings of this study are available from the corresponding
author upon reasonable request.

\begin{appendix}

\section{Floquet Theory}\label{Floquet theory}
Floquet theory was formulated by Gaston Floquet towards the end of the 19th century, in his attempt to solve linear differential equations with periodic coefficients  \citep{Floquet_ENS}.  Suppose one needs to solve the linear system, 
\beq
\dot{\bx} = \mathsfbi{A}(t)\bx\; , \quad \bx(t_0) = \bx_0,\label{eqn:linear_system}
\eeq
where $\mathsfbi{A}\in\mathcal{M}_{n\times n}$ is periodic in $t$ with period $T$. 
Considering $n$ linearly independent solutions $\{\bx_{1},\dots\bx_{n}\}$ 
of Eq.\eqref{eqn:linear_system}, it is useful to introduce the so-called fundamental matrix $\mathsfbi{X}$ by grouping the $\bx_i$ together
\beq
\mathsfbi{X}(t,t_0) := \Big( \bx_1;\bx_2;\dots ;\bx_n\Big),
\eeq
where the $\bx_i$ are column vectors, and the second argument has been added to specify that the initial condition occurs at time $t_0$. Thus, Eq.\eqref{eqn:linear_system} can be rewritten as 
\beq
\frac{d}{dt}\mathsfbi{X}(t,t_0) = \mathsfbi{A}(t)\mathsfbi{X}(t,t_0).\label{eqn:linear_fundamental}
\eeq 
If $\mathsfbi{X}(t_0,t_0) = \mathsfbi{I}$, $\mathsfbi{X}$ is called the \emph{principal} fundamental matrix. We now state some well-known theorems of Floquet theory, the proofs of which can be found in \citet{Meiss_linear}.\\

\begin{theorem}[Abel]
\label{Abel}
The determinant of the fundamental matrix $\mathsfbi{X}$ is 
\beq
\det \mathsfbi{X}(t,t_0) = \exp\int_{t_0}^{t}\text{tr}{\mathsfbi{A}(s)}ds. 
\eeq
Moreover, when $(t_0,t)=(0,T)$, it can be rewritten as the product of the so-called \emph{Floquet multipliers}
\beq
\det \mathsfbi{X}(T,0) = \prod_{i=1}^{n}e^{\nu_{i}T},
\eeq
where the $\nu_i$ are the Floquet exponents.  
\end{theorem}
\noindent The second statement implies that the Floquet multipliers are the eigenvalues of the monodromy matrix $\mathsfbi{m}:= \mathsfbi{X}(T,0)$, of the linear system Eq.\eqref{eqn:linear_fundamental}.\\

\begin{theorem}[Floquet-Lyapunov]
\label{Floquet-Lyapunov}
    The fundamental matrix $\mathsfbi{X}$ solution of the system Eq.\eqref{eqn:linear_fundamental} is of the form 
\beq
\mathsfbi{X}(t,t_0) = \mathsfbi{P}(t)e^{(t-t_0)\mathsfbi{B}}\label{eq:Floquet_thm}
\eeq
where the matrix $\mathsfbi{P}$ is symplectic and $T$-periodic, with $\mathsfbi{P}(t_0)=\mathsfbi{I}$ and $\mathsfbi{B}$ is a constant Hamiltonian matrix, that is
\beq
\mathsfbi{J}\mathsfbi{B} = \mathsfbi{B}^{T}\mathsfbi{J}^{T}, \quad \mathsfbi{J} = \left(\begin{array}{cc} \mathsfbi{0} & - \mathsfbi{I}   \\ \mathsfbi{I} & \mathsfbi{0}    \end{array}\right).
\eeq
\end{theorem}

\section{Frenet-Serret frame}\label{appendix:Frenet-Frame}

The Frenet-Serret frame is a local basis that spans the three-dimensional space $\mathbb{R}^3$. The terminology `local' arises from the fact that this frame is defined locally along a curve $\cC\subset \mathbb{R}^3$. Let us define the basis vectors, and some essential properties of the frame. We refer to \cite{Duignan_2021}.\\

\noindent The frame is composed of the three well-known tangent, normal and binormal vectors, defined and related to each-other as follows. Provided that the curve $\cC$ is described by the vector field $\br_0\in \mathbb{R}^3$ and parameterized by the arc length $\ell$, the tangent vector $\bt$ is defined as 
\beq
\bt := \frac{d\br_0}{d\ell} = \br_0'.
\eeq
The normal vector accounts for the normalized rate of change of the tangent along the curve:
\beq
\bn := \bt'\left| \bt'\right|^{-1}.
\eeq
Finally, the binormal vector is the cross product of the tangent and the normal vector:
\beq
\bb := \bt \times \bn.
\eeq
The curvature $\kappa$ and the torsion $\tau$ of the curve $\cC$ can be expressed from  the derivatives of $\br_0$: 
\beq
\kappa := |\br_0''|=|\bt'|, \quad \tau := \frac{\left( \br_0'\times \br_0''\right)\cdot \br_0'''}{\kappa^2},
\eeq
provided that the curvature is non-vanishing. The rate of change of the Frenet-Serret frame along the curve $\cC$ writes in terms of the curvature and the torsion. The resulting expressions are the so-called Frenet-Serret formulæ: 
\beq
\frac{d}{d\ell}\left(\begin{array}{c}\bt \\ \bn \\ \bb\end{array}\right) = \left(\begin{array}{ccc} 0 & \kappa(\ell)  & 0   \\ - \kappa(\ell) & 0  & \tau(\ell)  \\ 0  & - \tau(\ell)  & 0  \end{array}\right)\left(\begin{array}{c}\bt \\ \bn \\ \bb\end{array}\right).
\eeq
With the previous relations, the reader is equipped with what is necessary to delve into near-axis expansion coordinates, based on the Frenet-Serret frame.


\section{Near-axis expansion coordinates}\label{appendix:near-axis-coordinates}
This section is dedicated to the description of two set of coordinates that have been proven to be powerful in the expansion of operators or physical quantities in the vicinity of field lines. The terminology \emph{near-axis} expansion is used here, as in this paper, expansions are carried out around magnetic axes, but those coordinates remain suitable for any \emph{near-field line} expansion. Section (\ref{Appendix:mercier}) is dedicated to the description of Mercier coordinates, introduced by \citet{Mercier_1964}, and Section (\ref{Sol_Shaf_appendix}) deals with the Solov'ev-Shafranov coordinates, as described in \cite{PRL5}. They are both closely related, and a correspondence can easily be established between the two. In addition, they both are based on the Frenet-Serret frame, as described in Appendix (\ref{appendix:Frenet-Frame}). We emphasize that in what follows, the coordinates systems are introduced in the context of magnetic fields, and that each curve that is dealt with is assumed to be a field line. 

\subsection{Mercier coordinates } \label{Appendix:mercier}

\begin{figure}
\centering
\includegraphics[width = 0.5 \textwidth]{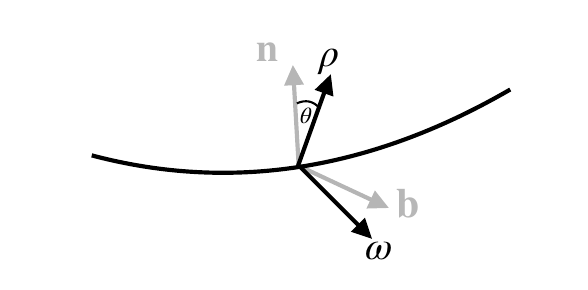}
\caption{Mercier's triad $\{\bt, \boldsymbol{\rho}, \boldsymbol{\omega} \}$ related to the usual Frenet-Serret $\{\bt, \boldsymbol{n}, \boldsymbol{b}\}$. The plain, black line is a field line.} \label{Fig:Mercier_coord}
\end{figure}

Mercier's coordinate system is based on the Frenet-Serret frame $\{\bt,\bn,\bb\}$, and related to the latter by a rotation of the normal and binormal vectors by a polar angle $\theta$, which is a purely geometric quantity, as shown in Fig.(\ref{Fig:Mercier_coord}).
The field-line (that can be considered to be a magnetic axis here) $\cC\subset \mathbb{R}^3$ and described by $\br_0$ is assumed closed and parameterized by the arc length $\ell$ with a total length $L$. 
Following \citet{MercierLuc}, the near-axis expansion is based on the construction of a tube of radius $\rho$ around the axis, such that any neighboring point can be described by  $\rho(\ell) \bsrho(\ell)$,
\begin{align}
    \br(\ell)= \br_0(\ell) + \rho(\ell) \bsrho(\ell), \quad \br_0'(\ell) =\bt(\ell),
    \label{eq:pos_vect_Mercier}
\end{align}
where the dependence in $\ell$ has been made explicit, but will be omitted in what follows for the sake of readability. 
However, it will be made clear whenever the dependence on quantities is not obvious. 
Therefore, $\{\bt,\bsrho,\bsomega\}$ is a right-handed triad related to  $\{\bt,\bn,\bb\}$ by 
\begin{align}
   \bsrho = \bn \cos{\theta} + \bb \sin{\theta} , \quad \bsomega = \bb \cos{\theta} -\bn \sin{\theta}. 
\end{align}

It can be showed by direct computation that $(\rho,\omega = \theta +\int \tau d\ell',\ell)$, with $\tau$ denoting the torsion of the axis $\br_0$ (see Appendix \ref{appendix:Frenet-Frame}), forms an orthogonal coordinate system with metric
\beq
    ds^2 = d\rho^2 + \rho^2 d\omega^2  +h^2 d\ell^2, \quad h= 1-\kappa \rho \cos{\theta},
    \label{eq:metric_Mercier}
\eeq
where $\kappa$ denotes the curvature of $\cC$. The product $\kappa\rho$ can be seen as a measure of the ratio between the radius of curvature of $\br_0$ and the radius of expansion around the latter. Therefore, from the previous characterization of the Mercier coordinates, a few differential identities can be derived. Since the basis considered for Mercier coordinates is the triad $\{\bt, \bsrho, \bsomega\}$, the identity tensor trivially writes 
\beq
\mathsfbi{I} =  \bt \bt+ \bsrho \bsrho+ \bsomega \bsomega. 
\eeq
By mean of the metric expression Eq.(\ref{eq:metric_Mercier}), the gradient can be derived:
\beq
\nabla = \bsrho \partial_\rho + \frac{\bsomega }{\rho}\partial_\omega +\frac{\bt}{h}\partial_\ell,\label{eq:gradient_mercier}
\eeq
where the subscripts denote with respect to which coordinate the partial derivative is taken. Moreover, some pseudo Poisson formulae can be derived for the derivatives of $\bsrho$ and $\boldsymbol{\omega}$, to read
\beq
\bsrho_{,\omega} = \bsomega, \quad \bsrho_{,\ell}= -\bt\; \kappa \cos{\theta}  , \quad  \bsomega_{,\omega} = -\bsrho, \quad \bsomega_{,\ell}= \bt\; \kappa \sin{\theta},
\eeq
where the comma represents partial differentiation. Finally, using Eq.(\ref{eq:gradient_mercier}), the gradient of each basis vector can be easily expressed: 
\beq
\nabla \bt =  \frac{\kappa}{h}\bt \bn, \quad 
    \nabla \bsrho= \frac{1}{\rho}\bsomega\bsomega - \frac{\kappa \cos{\theta}}{h}\bt \bt , \quad  \nabla \bsomega = \frac{\kappa \sin{\theta}}{h}\bt \bt
    -\frac{1}{\rho}\bsomega\bsrho.
\eeq
In the formalism developed by Mercier, when expanding quantities in powers of $\rho$, it can be useful to have an additional `phase' term $\delta$ that may simplify expressions. The latter phase is a function of the arc length $\ell$ and enables to define the so-called Mercier angle $u$:
\beq
u:=\theta + \delta = \omega-\int \tau d\ell' + \delta.
\eeq
In fact, the $\delta$ term describes the rotation of trajectories around the axis $\br_0$. One notes that by periodicity, $\delta$ has to satisfy $\delta(\ell + L) = \delta(\ell)+ 2\pi n$, with $n\in \mathbb{Z}$. The rotation function $\delta$ is important for the Solov'ev-Shafranov near-axis expansion as the latter is based on the construction of ellipses around $\br_0$, whose rotation is naturally important.

\subsection{Solov'ev-Shafranov coordinates } \label{Sol_Shaf_appendix}

The Solov'ev-Shafranov coordinates system is closely related to the Mercier triad, but differs in that the expansion is not carried by constructing a tube of radius $\rho$ around the axis, but an ellipse of semi-axes varying along $\br_0$. Let $\{\boldsymbol{t},\boldsymbol{\mathcal{N}},\boldsymbol{\mathcal{B}}\}$ be the orthogonal triad related to the Frenet-Serret frame through a rotation by function $\delta$ introduced in the Mercier formalism such that:
\beq
\boldsymbol{\mathcal{N}}= \boldsymbol{n}\cos\delta -\boldsymbol{b}\sin\delta = \bsrho\cos u -\bsomega\sin u, \quad \boldsymbol{\mathcal{B}}=\boldsymbol{n}\sin\delta +\boldsymbol{b}\cos\delta = \bsrho \sin u +\bsomega \cos u, \label{eq:cal_N_B}
\eeq
\begin{figure}
\centering
\includegraphics[width = 0.5 \textwidth]{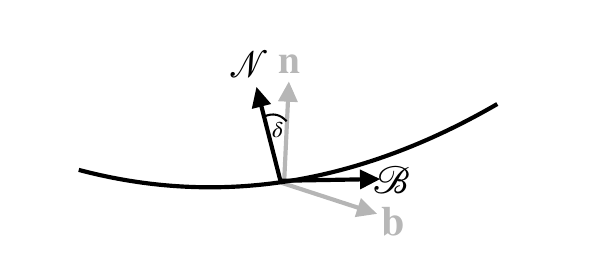}
\caption{Solov'ev-Shafranov triad $\{\bt, \boldsymbol{\mathcal{N}}, \boldsymbol{\mathcal{B}} \}$ related to the usual Frenet-Serret $\{\bt, \bn, \bb\}$. The plain, black line is a field line.} \label{Fig:Solov_shafran}
\end{figure} 
\noindent where $\{\bt, \bsrho, \bsomega\}$ form the Mercier basis, with Mercier angle $u$ as defined in Appendix \ref{Appendix:mercier}. Eq.(\ref{eq:cal_N_B}) shows how intrinsically related the two frames are.
Again, the field line considered, $\boldsymbol{r}_0$, forms a closed curve $\mathcal{C}$ parameterized by the arc length $\ell$ and a total length $L$. It is obvious that $\bcN$ and $\bcB$ depend on $\ell$, enabling to write 
\begin{align}
    \boldsymbol{\mathcal{N}}'= -\kappa \cos\delta\bt - u'\boldsymbol{\mathcal{B}}, \quad  \boldsymbol{\mathcal{B}}'=-\kappa \sin\delta \bt+ u' \boldsymbol{\mathcal{N}},
\end{align}
where $\kappa$ denotes the curvature of $\br_0$, which is a local property. Following \cite{PRL5}, the position vector $\boldsymbol{r}$ of any point in the vicinity of the axis can be expressed as
\begin{align}
    \boldsymbol{r}=\boldsymbol{r}_0 + x \boldsymbol{\mathcal{N}}+y \boldsymbol{\mathcal{B}},
    \label{eq:pos_vec_Sol_Shaf}
\end{align}
where $x$ and $y$ are expansion parameters depending on the position of the frame $\{\bt, \boldsymbol{\mathcal{N}}, \boldsymbol{\mathcal{B}} \}$ through $\ell$. In Mercier coordinates, $x=\rho \cos u$ and $y=\rho\sin u$. Differentiating $\br$ with respect to the length parameter $\ell$, 
\begin{align}
    \frac{d\boldsymbol{r}}{d\ell}= h \bt + (x'+u' y) \boldsymbol{\mathcal{N}} + (y'-u' x) \boldsymbol{\mathcal{B}},
    \label{eq:drdl_Sol_Shaf}
\end{align}
where the prime denotes total differentiation with respect to $\ell$, and $\br_0'=\boldsymbol{t}$. Additionally, $h=1-\kappa(x\cos \delta - y \sin \delta) = 1-\kappa\rho \cos \theta$, and $u'=\delta'-\tau$. It induces the following metric:
\beq
    ds^2=dx^2+dy^2+2 u' (y dx-x dy)d\ell + \left[h^2+{u'}^2(x^2+y^2) \right]d\ell^2.
\eeq
The covariant and contravariant forms $g_{ij}$ and $g^{ij}$ of the metric tensor for $ds^2$ read  
\beq
    g_{ij}=
    \begin{pmatrix}
        1 & 0 & u' y\\
        0 & 1 & -u' x\\
        u'y & \quad -u'x & \quad h^2+{u'}^2(x^2+y^2)
    \end{pmatrix}
    \label{eq:g_ijs}
\eeq
and 
\beq
     g^{ij}=\frac{1}{h^2}
    \begin{pmatrix}
        h^2+{u'}^2 y^2 &\quad  -{u'}^2 x y & \quad -u' y\\
        -{u'}^2 x y & 1 & u' x\\
        -u'y & \quad u'x & \quad 1
    \end{pmatrix}
    \label{eq:g^ijs}
\eeq
with $\sqrt{|g|}=h$. A direct computation shows that the following basis vectors lead to the same metric tensors:  
\begin{align*}
    \boldsymbol{e}^1 &= \nabla x=\boldsymbol{\mathcal{N}}-\frac{u' y}{h}\boldsymbol{t}, \quad &&\boldsymbol{e}^2= \nabla y=\boldsymbol{\mathcal{B}}+\frac{u' x}{h}\boldsymbol{t},  &&\boldsymbol{e}^3= \nabla \ell = \frac{1}{h}\boldsymbol{t},\\
    \boldsymbol{e}_1 &= h \left(\nabla y \times \nabla \ell\right) = \boldsymbol{\mathcal{N}},  \quad &&\boldsymbol{e}_2= h\left( \nabla \ell\times \nabla x\right)= \boldsymbol{\mathcal{B}},  &&\boldsymbol{e}_3= h\left(\nabla x \times \nabla y\right)\\
    & && && \hspace{11pt} = h \boldsymbol{t}+u'(-x\bcB + y \bcN),\label{eq:basis_vectors_sol_shaf}
\end{align*}
which will be useful and in fact, more convenient in the near-axis expansion of the quantities of interest in this paper. Here again, some expressions and differential identities can be derived in Solov'ev-Shafranov coordinates. The identity tensor writes
\beq
\mathsfbi{I} =  \bt \bt+ \bcN\bcN+ \bcB\bcB, 
\eeq
or alternatively 
\beq
\mathsfbi{I} =  \boldsymbol{e}_1\boldsymbol{e}_1 + \boldsymbol{e}_2\boldsymbol{e}_2 + \boldsymbol{e}_3\boldsymbol{e}_3. 
\eeq
The gradient in the basis $\{ \boldsymbol{e}_j\}_j$ is expressed as 
\beq
\grad \equiv \mathbf{e}_j\; g^{ij}\frac{\partial}{\partial \alpha^{i}}\label{eq:grad_solShaf}
\eeq
with $\alpha^1=x$, $\alpha^2=y$, $\alpha^3=\ell$, and the Einstein's summation convention.


\section{Derivation of the rotational transform from the action in Mercier's coordinates}\label{app:mercier_derivation}

\subsection{Near-axis expansion of the second variation tensor}

 As we did in the Solov'ev-Shafranov geometry in the body of this paper, in order to expand the operator 
 \begin{align}
    \mathsfbi{M} :=  \frac{\delta^2 \mathcal{S}}{\delta \bx \delta \bx}= -\left(\mathsfbi{I}\times \bB \right) \frac{d}{d\ell}+ \bx' \times (\nabla \bB)^\mathsf{T},
    \label{eq:TensorM_appendix}
\end{align}
 in Mercier's coordinates,  $\bB$ is expanded in the  parameter $\rho$ such that $\kappa \rho \ll 1$. To evaluate $\mathsfbi{M}$ to lowest order in $\rho$, the magnetic field needs to be expanded up to first order. Therefore, we set
\beq
    \bB = B_0(\ell) \bt + \rho \bB_1 , \quad \bB_1= \left( B_{1}^{\rho} \bsrho +  B_{1}^{\omega}\bsomega + B_{1}^{t}\bt \right).
    \label{eq:form_B}
\eeq
The next step in the expansion of $\mathsfbi{M}$ is to compute $\grad \bB$. It is straightforward from the differential identity Eq.\eqref{eq:gradient_mercier}:
\begin{align}
    \nabla \bB = \frac{1}{h} \left(\bt \bt \;  B_0'(\ell) + \bt \bn\; \kappa B_0  \right) + \bsrho \; \bB_1 + \rho \nabla \bB,
\end{align}
such that, to leading order
\beq
\begin{split}
    \bx'\times \nabla \bB^\mathsf{T} &=\kappa B_0 \bb \bt + \left(B_{1}^{\rho} \bsomega  -B_{1}^{\omega} \bsrho \right)\bsrho + \bsomega\bsomega\;(\partial_\omega B_{1}^{\rho}-B_{1}^{\omega})  -\bsomega\bsrho\; (\partial_\omega B_{1}^{\omega}+B_{1}^{\rho}),\\
    -(\mathsfbi{I}\times \bB) &=(\bsrho\bsomega -\bsomega\bsrho)B_0.
\end{split}\label{eq:GradB_tensor}
\eeq
Finally, $\mathsfbi{M}$ as given by Eq.\eqref{eq:TensorM} simplifies to
\beq
\begin{split}
    \mathsfbi{M} & := \mathsfbi{M}_1 \frac{d}{d\ell}+\mathsfbi{M}_2, \\
    \mathsfbi{M}_1 &= (\bsrho\bsomega -\bsomega\bsrho)B_0 ,\\
    \mathsfbi{M}_2 &=  \kappa B_0 \bb \bt + \left(B_{1}^{\rho} \bsomega  -B_{1}^{\omega} \bsrho \right)\bsrho + \bsomega\bsomega\;(\partial_\omega B_{1}^{\rho}-B_{1}^{\omega})  -\bsomega\bsrho\; (\partial_\omega B_{1}^{\omega}+B_{1}^{\rho}).\label{eq;Tensor_M_exp}
\end{split}
\eeq
So far, the power expansion of the gradient of the magnetic field relied on a generic expression of $\bB$ in terms of the $B_{1}^{\rho}$, $B_{1}^{\omega}$ and $B_{1}^{t}$. The latter functions can be determined enforcing additional constraints such as $\Div \bB=0$ and $\rot \bB = \boldsymbol{J}_0=J_0(\ell)\bt$. The divergence and curl of $\bB$ can be evaluated directly from the $\nabla \bB$ tensor, using dyadic algebra. From
\beq
\begin{split}
    \nabla \cdot \bB = \mathsfbi{I}\boldsymbol{:}\nabla \bB &= B_0' + 2 B_{1}^{\rho} + \partial_\omega B_{1}^{\omega}\\
    \nabla \times \bB = -\mathsfbi{I}\boldsymbol{\overset{\times}{.}}\nabla \bB &= \bt\; (2 B_{1}^{\omega}-\partial_\omega B_{1}^{\rho})  + \left(\bsomega -\bsrho\partial_\omega\right)(B_0 \kappa \cos\theta -B_{1}^{t}),
\end{split}
\eeq
where $\partial_{\omega}:=\partial/\partial \omega$, it can be easily shown that
\begin{align}
    B_{1}^{t}= \kappa B_0 \cos\theta=B_{1t}, \quad B_{1}^{\rho}= -\frac{1}{2}(B_0' + \partial_\omega b_{1}) = B_{1\rho}, \quad B_{1}^{\omega}= \frac{1}{2}J_0+b_1=B_{1\omega},
    \label{eq:B1x_exp}
\end{align}
where, $b_1$ satisfies the Laplace equation
\beq
    \left(\partial_\omega^2 +4\right)b_1=0.
    \label{eq:laplace_b1}
\eeq
 Following \cite{MercierLuc}, the solution of Eq.\eqref{eq:laplace_b1} can be expressed as
 \beq
    \frac{b_1}{B_0}=b_{c2}\cos{(2u)} + b_{s2}\sin{(2u)} =  \tanh{\eta(\ell)}\left(\delta'-\tau +\frac{J_0}{2B_0}\right)\cos{(2u)}+\frac{\eta'}{2}\sin{(2u)},\label{eq:b1}
\eeq
where $u$ is the Mercier angle as introduced in Appendix (\ref{Appendix:mercier}). The functions $\eta(\ell)$ and $\delta(\ell)$ represent respectively the eccentricity and the rotation of the elliptic flux-surfaces winding around the axis, given by the flux function 
\beq
    \psi=\rho^2 B_0\left[ \cosh({\eta})+\sinh({\eta})\sin{(2u)} \right].
    \label{eq:psi_exp_Mercier}
\eeq

\subsection{Floquet exponents and rotational transform}

Now that the MHD constraints have been enforced, let us obtain the null eigenvector $\bv= v^t \bt + v^{\rho} \bsrho+ v^{\omega} \bsomega$ of $ \mathsfbi{M}$ such that $ \mathsfbi{M}\bv = \mathbf{0}$, in order to determine the Floquet exponents. We take advantage of the fact that Mercier's coordinates are orthogonal, so the contravariant and covariant coordinates of a vector in that basis are equal, to avoid a heavy notation, with superscript for components \emph{and} derivatives. However, when doing similar derivations, one has to be careful with the nature of the components they are working with. We state that the components $v_t,v_\rho, v_\omega$ are functions of $\omega$ and $\ell$. The $\mathsfbi{M}_1$ terms can be shown to be

\begin{align}
   \mathsfbi{M}_1 \frac{d}{d\ell} \bv = B_0 \left[ \bsrho\;(v_\omega'-\kappa \sin\theta\; v_t)-\bsomega (v_\rho' +\kappa \cos\theta\; v_t) \right].
    \label{eq:M1_v}
\end{align}
Similarly,
\beq
\begin{split}
     \mathsfbi{M}_2 \bv =&\; \bsrho\;\left[B_0 v_\omega' +v_\omega (B_0'+B_{1\rho})-B_{1\omega}v_\rho\right]\\
     &+\bsomega\;\left[-B_0 v_\rho' +v_\omega (B_{1\omega}-J_0)+B_{1\rho}v_\rho\right].
\end{split}
\eeq
Thus, we observe that $\mathsfbi{M}\bv =\mathbf{0}$ only has 
$\bsrho,\bsomega$ components with $\ell$ derivatives of $v_\rho,v_\omega$. The tangential component $v_t$ is thus an arbitrary constant, which can be absorbed by redefining $v_\rho,v_\omega$. We shall therefore set $v_t=0$. The condition $\mathsfbi{M}\bv =\mathbf{0}$ leads to the system
\begin{align}
    B_0 
    \begin{pmatrix}
     \;\;0 & +1 \;\; \\-1 & 0 
    \end{pmatrix}
    \frac{d}{d \ell} 
    \begin{pmatrix}
        v_\rho\\ v_\omega
    \end{pmatrix} +
    \begin{pmatrix}
        B_{1\omega} & & [B_0'+B_{1\rho}]\\ B_{1\rho} & & [B_{1\omega}-J_0]
    \end{pmatrix}
     \begin{pmatrix}
        v_\rho\\ v_\omega
    \end{pmatrix} =0.
    \label{eq:Mdotv_matrix_form}
\end{align}
Substituting the constraints Eq.(\ref{eq:B1x_exp}) into Eq.\eqref{eq:Mdotv_matrix_form}, we obtain the equivalent system of coupled linear PDEs
\begin{equation}
\left\{ \begin{array}{l}
\displaystyle
v_\omega' +\left( \frac{B_0'}{2B_0}-\frac{\partial_\omega b_1}{2B_0} \right) v_\omega -\left(\frac{J_0/2}{B_0}+\frac{b_1}{B_0}\right) v_\rho = 0 \\[16pt]
\displaystyle
    v_\rho' +\left( \frac{B_0'}{2B_0}+\frac{\partial_\omega b_1}{2B_0} \right) v_\omega +\left(\frac{J_0/2}{B_0}-\frac{b_1}{B_0}\right) v_\omega   = 0. 
\end{array}\right.\label{eq:Mdotv_PDE_system}
\end{equation}
Therefore, the the system Eq.\eqref{eq:Mdotv_PDE_system} can be represented in the form 
\begin{align}
    \frac{d\tilde{\mathbf{v}}}{d\ell}=\mathsfbi{A}(\ell) \tilde{\mathbf{v}}, \quad \tilde{\mathbf{v}}= (v_\rho, v_\omega)^{\mathsf{T}}.
    \label{eq:equiv_matrix_form}
\end{align}
Here, $\mathsfbi{A}$ is a matrix with components periodic in $\ell$ with period $L$. Note that this representation is possible since all the quantities between brackets are depending on $\ell$ along the field line from which the expansion is carried, $\omega$ included \citep{PRL5, Duignan_2021}. We can directly use Floquet theorem on Eq.\eqref{eq:equiv_matrix_form} to conclude that the solution must be of the form
\beq
    \tilde{\bv}=\mathsfbi{U}(\ell)e^{\mathsfbi{C} \ell/L},
    \label{eq:Floquet_theorem}
\eeq
where $\mathsfbi{U}$ is a symplectic periodic matrix and $\mathsfbi{C}$ is a constant Hamiltonian matrix. The eigenvalues of $\mathsfbi{C}$ are the Floquet exponents, which must be of the form $\pm i \nu$ ($\nu \in \mathbb{R}$) near an elliptic axis.\\

An equivalent approach to solve the system from Eq.\eqref{eq:Mdotv_PDE_system} and hence identify the exponents $\nu$, is to use the fact that $b_1$ only has second harmonics in $u$. This allows us to seek a solution, where $\tilde{\bv}$ only has first harmonics in $u$. From now on, we will follow this approach. We will show that there exist  solutions to $\mathsfbi{M}\bv =\mathbf{0}$ with
\begin{align}
    \begin{pmatrix}
        v_\rho(\ell) \\ v_\omega(\ell)
    \end{pmatrix}
    = \begin{pmatrix}
        v_{\rho c}(\ell) \\ v_{\omega c}(\ell)
    \end{pmatrix} \cos u
    + \begin{pmatrix}
        v_{\rho s}(\ell) \\ v_{\omega s}(\ell)
    \end{pmatrix}\sin u.
    \label{eq:v_sin_cos_form}
\end{align}
Substituting the above into Eq.\eqref{eq:Mdotv_PDE_system}, together with the expressions for $B_{1\rho}, B_{1\omega}$ from Eq.\eqref{eq:B1x_exp} yields first and third order harmonics in $u$. Equating the terms with first harmonics in $u$, the $v_\omega'$ equation yields
\begin{equation}
\left\{ \begin{array}{l}
\displaystyle
v_{\omega c}' -(\tau -\delta')v_{\omega s} -\left(\frac{J_0/2}{B_0}\right) v_{\rho c} + \frac{B_0'}{2 B_0} v_{\omega c} + \frac{1}{2}(\mathfrak{a}_{c2c} + \mathfrak{a}_{s2s} ) =0 \\[16pt]
\displaystyle
    v_{\omega s}' +(\tau -\delta')v_{\omega c} -\left(\frac{J_0/2}{B_0}\right) v_{\rho s} + \frac{B_0'}{2 B_0} v_{\omega s} + \frac{1}{2}(\mathfrak{a}_{s2c} - \mathfrak{a}_{c2s} ) =0. 
\end{array}\right.\label{eq:v_omega_H1}
\end{equation}
Equating the third harmonics terms lead to the following constraints
\begin{align}
\mathfrak{a}_{c2c}= \mathfrak{a}_{s2s} , \quad \mathfrak{a}_{s2c}=- \mathfrak{a}_{c2s},
    \label{eq:v_omega_H3}
\end{align}
where the $\mathfrak{a}$ terms have the following definitions 
\beq
\begin{split}
    \mathfrak{a}_{c2c}&= -\frac{b_{c2}}{B_0}v_{\rho c}-\frac{b_{s2}}{B_0}v_{\omega c}, \quad \mathfrak{a}_{s2s}= -\frac{b_{s2}}{B_0}v_{\rho s}+\frac{b_{c2}}{B_0}v_{\omega s}, \label{eq:mfas}\\
    \mathfrak{a}_{s2c}&= -\frac{b_{s2}}{B_0}v_{\rho c}+\frac{b_{c2}}{B_0}v_{\omega c}, \quad \mathfrak{a}_{c2s}= -\frac{b_{c2}}{B_0}v_{\rho s}-\frac{b_{s2}}{B_0}v_{\omega s}. 
\end{split}
\eeq
The constraint Eq.\eqref{eq:v_omega_H3} and the definitions Eq.\eqref{eq:mfas} imply, together with $b_{s2}^2+b_{c2}^2\neq 0$ that
\begin{align}
    v_{\rho s}= v_{\omega c}, \quad v_{\rho c}=- v_{\omega s},
     \label{eq:reln_v_rho_v_omega}
\end{align}
which allows us to rewrite Eq.\eqref{eq:v_omega_H1} solely in terms of $v_{\omega c}, v_{\omega s}$. Simplification leads to
\begin{equation}
\hspace{-20pt}\left\{ \begin{array}{l}
\displaystyle
v_{\omega c}' + \left(\frac{B_0'}{2 B_0} -\frac{b_{s2}}{B_0} \right) v_{\omega c} +\left( \frac{J_0/2}{B_0}- \tau +\delta' +\frac{b_{c2}}{B_0}\right) v_{\omega s}  =0 \\[16pt]
\displaystyle
v_{\omega s}' + \left(\frac{B_0'}{2 B_0} +\frac{b_{s2}}{B_0} \right) v_{\omega s} -\left( \frac{J_0/2}{B_0}- \tau +\delta' -\frac{b_{c2}}{B_0}\right) v_{\omega c} =0. 
\end{array}\right.\label{eq:v_omega_simpl}
\end{equation}
Finally, substituting Eq.\eqref{eq:b1}, we get
\begin{equation}
\hspace{10pt}\left\{ \begin{array}{l}
\displaystyle
   v_{\omega c}' + \left(\frac{B_0'}{2 B_0} -\frac{\eta'}{2} \right) v_{\omega c} +\frac{e^{+\eta}}{\cosh{(\eta)}}\left( \frac{J_0/2}{B_0}- \tau +\delta' \right) v_{\omega s}  =0  \\[16pt]
\displaystyle
   v_{\omega s}' + \left(\frac{B_0'}{2 B_0} +\frac{\eta'}{2}  \right) v_{\omega s} -\frac{e^{-\eta}}{\cosh{(\eta)}}\left( \frac{J_0/2}{B_0}- \tau +\delta'\right) v_{\omega c} =0. 
\end{array}\right.\label{eq:v_omega_eqns}
\end{equation}
It is possible to further simplify the system by introducing new variables $X,Y$,
\begin{align}
    v_{\omega c}= \frac{1}{\sqrt{B_0'}}e^{+\eta/2}X(\ell), \quad  v_{\omega s}= \frac{1}{\sqrt{B_0'}}e^{-\eta/2}Y(\ell)
\end{align}
\noindent such that Eq.\eqref{eq:v_omega_eqns} reduces to
\begin{align}
    X' + \Omega_0(\ell) Y =0, \quad Y'-\Omega_0(\ell) X =0, \quad \Omega_0(\ell) = \frac{\frac{J_0/2}{B_0}- \tau +\delta'}{ \cosh{\eta}}.
    \label{eq:X_Y_system}
\end{align}
One notes that Eq.\eqref{eq:X_Y_system} has the exact same form as Eq.(\ref{eq:oscill_harm_SolShaf}), which describe a harmonic oscillator system with a `time'-dependent frequency $\Omega_0$. Once again, using the complex variable $Z=X+i Y$, the system can be rewritten as a single complex ODE
\begin{align}
    Z'-i \Omega_0 Z =0 \quad \Rightarrow \quad Z(\ell)= Z_0 \exp{\int_0^\ell i\Omega_0(s) ds}.
\end{align}
Separating the periodic and non-periodic parts of the exponential we get
\begin{align}
    Z(\ell) = Z_p(\ell) e^{i \nu \ell/L}, \quad Z_p(\ell)= Z_0 \exp{\int_0^\ell i\tilde{\Omega}_0(s) ds}, \label{eq:Z_expr}\\
    \overline{\Omega} \equiv \frac{1}{L}\int_0^L \Omega_0(s) ds, \quad \tilde{\Omega_0}= \Omega_0(\ell) - \overline{\Omega}.
    \nonumber
\end{align}
Comparing Eq.\eqref{eq:Z_expr} with Eq.\eqref{eq:Floquet_theorem}, we find that $\nu$, given by 
\begin{align}
\nu = \int_0^L \Omega_0(s)\; ds \pmod{2\pi} = \oint \frac{\frac{J_0(s)/2}{B_0(s)}- \tau(s) +\delta'(s)}{\cosh{\eta(s)}}\; ds \pmod{2\pi},
    \label{eq:iotabarinMercier}
\end{align}
is the Floquet exponent for the system. It confirms the result obtained in the Solov'ev-Shafranov coordinates, and it is a second confirmation that the second-variation of the magnetic field action, when expanded by mean of a near-axis formalism and combined with Floquet theory, yields the correct on-axis rotational transform. This was expected since the result is coordinate independent.


\section{Discrete formalism - the piecewise linear discretization}\label{app:piecewise}

A discrete method to compute the on-axis Floquet exponent has been introduced in Section  \ref{sec:discrete_action}. Here, we give the computations in the particular case of the curve is broken down to a concatenation of segments. Recall that the discrete action was introduced as
\beq
\mathcal{S} = \sum_{i=1}^{n-1} \int_{\mathcal{C}_i} \bA \cdot d\boldsymbol{\ell},
\label{eq:discrete_action_appendix}
\eeq
 Thus, taking the $\mathcal{C}_i$ to be segments, the action sums up to a sum of integral along piecewise linears $\mathcal{C}_i:=\{ \mathbb{R}^3\ni\bx = \zeta(\bx_{i+1}-\bx_{i}) + \bx_i \;|\; \zeta \in [0,1] \}$:
\beq
 \begin{split}
 \mathcal{S} &= \sum_{i=1}^{n-1} S(\bx_{i},\bx_{i+1}),\\
 S(\bx_{i},\bx_{i+1}) &= \int_0^{1} d\zeta\;\bA\Big(\zeta(\bx_{i+1}-\bx_i)+\bx_i\Big)\cdot (\bx_{i+1}-\bx_i)\\
 &=\int_0^{1}d\zeta\; \mathbf{A}\big(\bv(\bx(\zeta))\big) \cdot \bu,\\
 \end{split}
\eeq
\noindent with $\bx(0) = \bx_i$ and $\bx(1)=\bx_{i+1}$, and the vector fields $\bu$ and $\bv$ defined as follows:
\beq
\begin{split}
\bu(\bx_i,\bx_{i+1}) &= \bx_{i+1}-\bx_i,\\
\bv(\zeta,\bx_i, \bx_{i+1}) &= \zeta(\bx_{i+1}-\bx_i)+\bx_i, 
\end{split}
\eeq
such that $\nabla_{\bx_i} \bu(\bx_i,\bx_{i+1}) = -\mathsfbi{I}$, $\nabla_{\bx_{i+1}} \bu(\bx_i,\bx_{i+1}) = \mathsfbi{I}$,  $\nabla_{\bx_i}\bv(\zeta,\bx_i, \bx_{i+1}) = (1-\zeta) \mathsfbi{I}$, $\nabla_{\bx_{i+1}}\bv(\zeta,\bx_i, \bx_{i+1}) = \zeta\mathsfbi{I}$.
Since
\beq
\begin{split}
\nabla_{\bx_i} \Big[ \mathbf{A}(\bv(\bx_i)) \cdot\bu(\bx_i)\Big] &= \mathsfbi{J}^{\mathsf{T}}_{\mathbf{A} \circ \bv}(\bx_i) \cdot \bu(\bx_i) + \mathsfbi{J}^{\mathsf{T}}_{\bu}(\bx_i)\cdot \mathbf{A}(\bv(\bx_i))\\
&= \Big[ \mathsfbi{J}^{\mathsf{T}}_{\bv}(\bx_i)\cdot \mathsfbi{J}^{\mathsf{T}}_{\mathbf{A}}(\bv(\bx_i))\Big]\cdot \bu(\bx_i)+ \mathsfbi{J}^{\mathsf{T}}_{\bu}(\bx_i)\cdot \mathbf{A}(\bv(\bx_i)),\\
\end{split}
\eeq
where $\mathsfbi{J}_{\boldsymbol{a}}^{\mathsf{T}}$ stands for the transpose of the Jacobian matrix of the vector field $\boldsymbol{a}$, one gets the following first order derivatives of the action
\beq
\begin{split}
\mathbf{S}_1^{[i,i+1]} &= \int_0^{1}(1-\zeta)\mathsfbi{J}_{\bA}^{\mathsf{T}}\Big(\zeta(\bx_{i+1}-\bx_i)+ \bx_i\Big)\cdot(\bx_{i+1}-\bx_i) - \bA\Big(\zeta(\bx_{i+1}-\bx_i)+ \bx_i\Big) d\zeta,\\
\mathbf{S}_2^{[i-1,i]} &= \int_0^{1}\zeta \mathsfbi{J}_{\bA}^{\mathsf{T}}\Big(\zeta(\bx_{i}-\bx_{i-1})+ \bx_{i-1}\Big)\cdot(\bx_{i}-\bx_{i-1}) + \bA\Big(\zeta(\bx_{i}-\bx_{i-1})+ \bx_{i-1}\Big) d\zeta.
\end{split}
\eeq
A direct computation of the second derivatives reads
\beq
\begin{split}
\mathsfbi{S}_{12}^{[i-1,i]}&=\int_0^{1} d\zeta \zeta\Big[ (1-\zeta) \nabla_{\bv} \mathsfbi{J}^{\mathsf{T}}_{\bA}(\bv)\cdot (\bx_i - \bx_{i-1}) - \mathsfbi{J}^{\mathsf{T}}_{\bA}(\bv)\Big]+ (1-\zeta)\mathsfbi{J}_{\bA}(\bv), \\
\mathsfbi{S}_{21}^{[i,i+1]}&=\int_0^{1} d\zeta (1-\zeta)\Big[ \zeta \nabla_{\bv} \mathsfbi{J}^T_{\bA}(\bv)\cdot (\bx_{i+1} - \bx_{i}) + \mathsfbi{J}^T_{\bA}(\bv)\Big]+ \zeta \mathsfbi{J}_{\bA}(\bv),\\
\mathsfbi{S}_{22}^{[i-1,i]}&=\int_0^{1} d\zeta \zeta\Big[ \zeta \nabla_{\bv} \mathsfbi{J}^{\mathsf{T}}_{\bA}(\bv)\cdot (\bx_i - \bx_{i-1}) + \big(\mathsfbi{J}_{\bA}(\bv) +\mathsfbi{J}^{\mathsf{T}}_{\bA}(\bv)\big)\Big],\\
\mathsfbi{S}_{11}^{[i,i+1]}&=\int_0^{1} d\zeta (1-\zeta)\Big[ (1-\zeta) \nabla_{\bv} \mathsfbi{J}^{\mathsf{T}}_{\bA}(\bv)\cdot (\bx_{i+1} - \bx_{i}) - \big(\mathsfbi{J}_{\bA}(\bv)+\mathsfbi{J}_{\bA}^{\mathsf{T}}(\bv)\big)\Big].\\
\end{split}\label{eq:derivatives_linear}
\eeq
The second order derivatives above can be easily computed numerically. We remind the multipliers $\lambda$ of the curve are given by solutions of 
\beq
\begin{split}
\det \mathsfbi{M}(\lambda) &= (-1)^n \lambda ^{-3}\det\Big(\mathsfbi{T}_S-\lambda \mathsfbi{I}_{6} \Big)\det\Big( \prod_{i=1}^{n}\mathsfbi{S}_{12}[i,i+1]\Big)\\
\mathsfbi{T}_S &= \prod_{i=1}^{n}\left(\begin{array}{cc} -{\mathsfbi{S}^{-1}_{12}}^{[i,i+1]}(\mathsfbi{S}_{22}^{[i-1,i]}+\mathsfbi{S}_{11}^{[i,i+1]}) & \hspace{0.2cm}-{\mathsfbi{S}_{12}^{-1}}^{[i,i+1]}\mathsfbi{S}_{12}^{[i-1,i]}  \\ \mathsfbi{I}_{3} & \hspace{0.2cm} 0\end{array}\right),\\
\end{split}\label{eq:eigenvalues_transfer_app}
\eeq
and that they are linked to the Floquet exponent by $\lambda = e^{i\nu}$. From the set of expressions Eq.\eqref{eq:derivatives_linear}, the determinant of $\mathsfbi{M}$ can be computed, leading to the multipliers. The above can be implemented numerically as a novel tool to compute the on-axis Floquet-exponent. 

\end{appendix}

\bibliographystyle{tail/jpp}
\bibliography{tail/references}

\end{document}